\documentclass[preprint]{revtex4}
\usepackage{epsfig}

\usepackage{amsmath}
\usepackage{graphicx}
\usepackage{float}
\usepackage{subfigure}

\def\beq{\begin{equation}}
\def\eeq{\end{equation}}
\def\beqa{\begin{eqnarray}}
\def\eeqa{\end{eqnarray}}

\begin{document}

\title{\sc Sound waves and solitons in hot and  dense nuclear matter}

\author{D.A. Foga\c{c}a\dag\, L.G. Ferreira Filho\ddag\   and F.S. Navarra\dag\ }
\address{\dag\ Instituto de F\'{\i}sica, Universidade de S\~{a}o Paulo\\
 C.P. 66318,  05315-970 S\~{a}o Paulo, SP, Brazil}
\address{\ddag\ Faculdade de Tecnologia, Universidade do Estado do Rio de Janeiro \\
Via Dutra km 298, CEP 27523-000, Resende, RJ, Brazil}

\begin{abstract}

Assuming that  nuclear matter can be treated as a perfect fluid,  we study the propagation 
of perturbations in the baryon density. The equation of state is derived from a relativistic  
mean field model, which is a variant of the non-linear Walecka model. The expansion of the 
Euler and continuity equations of relativistic hydrodynamics around equilibrium 
configurations  leads to differential equations for the density perturbation.  
We solve them numerically for linear and spherical 
perturbations and follow the propagation of the initial pulses. For linear perturbations we
find single soliton solutions and solutions with one or more solitons followed by  
``radiation''. Depending on the equation of state a strong damping may occur. Spherical 
perturbations are strongly damped and almost do not propagate. We study these equations  
for matter at finite temperature.

\end{abstract} 

\maketitle



\vspace{1cm}
\section{Introduction}

Over the last decades hydrodynamics of strongly interacting systems 
\cite{hidro1,hidro2,hidro3,hidro4} has been applied to 
cold nuclear physics, to low 
and high energy nuclear reactions and to phenonema taking place in dense stars. Recently 
hydrodynamical models became more sophysticated and received more support from 
experimental data, in particular from the measurement of elliptic flow at RHIC \cite{v2}.  
Whereas other approaches can give  satisfactory descriptions of the measured  transverse 
momentum distributions, when it comes to elliptic flow, there are not many options other than 
hydrodynamical models.  There is compelling evidence that we have seen ``the perefect fluid'' 
at RHIC. This evidence might be significantly reinforced by the observation of waves. 
Waves in a hadronic medium are produced in many physical situations. In fact, there are already  
some indications that these waves have been formed.  
In relativistic heavy ion collisions  we may have hard parton - parton collisions in which the 
outcoming partons have to traverse the surrounding  fluid to escape and form jets. Their passage 
may form Mach shock waves \cite{shock}, which will affect the transverse momentum distribution 
of the  observed final particles. These ``Mach cones'' may have been observed  at 
RHIC \cite{away1,away2}. 

Under certain conditions  waves may form solitons. Therefore 
we can go a step further and look for solitons in a  hadronic medium. In the RHIC scenario, 
for example, the same supersonic motion that generates conical shock waves may also 
generate solitons. Whether or not this happens, depends on details of the equation of state 
and on the approximations used in the hydrodynamical description of the motion. 
Another scenario where solitons may appear is in the core of dense stars. Here perturbations in 
the baryon density may be caused, for example, by interactions of neutrinos with the baryonic  
matter. In a pioneering  series of works  on soliton formation in nuclear matter  
\cite{frsw,rww,abu} it was suggested that in 
nucleon  - nucleus collision at intermediate bombarding energies 
($\simeq 50 \, - \, 200$ MeV) the nucleon may be absorbed by the nucleus (treated as a 
fluid at rest) and propagate as a localized pulse of baryon density.

In this work 
we study the propagation of sound waves in dense and hot hadronic matter. More specifically 
we consider  the propagation of perturbations in the baryon density. These perturbations may 
generate ordinary waves, shock waves and also  Korteweg - de Vries (KdV) solitons.  Starting 
from the equations of relativistic hydrodynamics  at zero and finite temperature and  in 
Cartesian $(x,t)$  and 
spherical $(r,\theta,\phi,t)$ coordinates, we derive differential equations and find their 
numerical solutions. 
The equation of state is derived from a relativistic mean field model of the Walecka type 
\cite{lala, furn, serot}.  
We discuss the features of the solutions and the role played by the microscopic interactions 
in the shape and propagation of the sound waves.

In a  previous work \cite{fn1} we have studied the 
formation and propagation of KdV solitons in cold 
nuclear matter. We found 
that these solitary waves can indeed  exist in the nuclear medium, provided that derivative 
couplings between the nucleon and the vector field are included in the interaction 
Lagrangian. For this class of 
equation of state (EOS), which is quite general, perturbations 
on the nuclear density can propagate as a pulse without dissipation.  

During the analysis of 
several realistic nuclear equations of state, we realized that, very often the speed of 
sound $c_s$ is in the range $0.15 - 0.25$. Compared to the speed of light these values are 
not large but not very small either. This suggests that, even for slowly moving nuclear 
matter, relativistic effects might be sizeable. We investigated these effects in 
\cite{fn2} and in \cite{fn3}.

The propagation of density pulses might be relevant for the astrophysics of dense stars. This
motivated us to extend  our  results to the spherical geometry. 
In \cite{fn4} combining the Euler and continuity equations in relativistic 
hydrodinamics in  spherical coordinates, we have  obtained for the first time  an equation 
similar to the  KdV equation.  
Spherical KdV - like equations have been found before in other contexts, as, for example, 
in the study of nonlinear waves in dusty plasmas 
\cite{xue,pu}. 

In the present work we reexamine all our previous works, looking for the numerical solutions  
of the previously encountered differential equations. In the linear case we compare the 
analytical solution with the numerical one and study the sensitivity of the solution to  
the initial conditions. In the spherical case there is no analytical solution. Our 
numerical solution could be compared to the one found in \cite{xue,pu}. We 
have extended the formalism to finite temperature and, with the new equation of state, derived 
differential equations which are temperature dependent. We have also studied the limiting case 
where the differential equations generate shock waves.

The text is organized as folows. In  section II, for convenience, we collect some useful 
formulas for  hydrodyamics. In section III we present the equation of state obtained with
our model. In sections IV  we derive the  spherical  KdV-like equations and 
in section V we present and discuss the numerical solutions of these equations. Finally, 
in  section VI we present some conclusions.


\section{Relativistic hydroynamics}

In this section we review the main equations of  relativistic 
hydrodynamics. In natural units ($c=1$) the velocity four vector  $u^{\nu}$ is defined as:
\begin{equation}
u^{\nu} \,\, = \,\, (u^{0},\vec{u}) \,\,  = \left(\gamma , \gamma \vec{v} \right)
\label{us}
\end{equation}
where $\gamma$ is the Lorentz contraction factor given by $\gamma=(1-v^{2})^{-1/2}$.  
The velocity field of  matter is $\vec{v}=\vec{v}(t,x,y,z)$
and thus $u^{\nu}u_{\nu}=1$. The energy-momentum tensor is, as usual, given by:
\begin{equation}
T_{\mu \nu}=(\varepsilon +p)u_{\mu}u_{\nu}-pg_{\mu\nu}
\label{tensor}
\end{equation}
where  $\varepsilon$ and $p$ are the energy density and pressure respectively
($g_{00}= - g_{ii} = 1$ and $g_{\mu \nu} =0$ if $\mu \ne \nu$). 
Energy-momentum conservation is  ensured by:
\begin{equation}
\partial_{\nu}{T_{\mu}}^{\nu}=0
\label{cons}
\end{equation}
The projection of (\ref{cons}) onto  a direction perpendicular to $u^{\mu}$ gives us
the relativistic version of  Euler equation \cite{wein,land,elze}:
\begin{equation}
{\frac{\partial {\vec{v}}}{\partial t}}+(\vec{v} \cdot \vec{\nabla})\vec{v}=
-{\frac{1}{(\varepsilon + p)\gamma^{2}}}
\bigg({\vec{\nabla} p +\vec{v} {\frac{\partial p}{\partial t}}}\bigg)
\label{eul}
\end{equation}
The  continuity equation for the baryon number is  
\cite{wein,land,elze}:
\begin{equation}
\partial_{\nu}{j_{B}}^{\nu}=0
\label{conucleon}
\end{equation}
Since ${j_{B}}^{\nu}=u^{\nu} \rho_{B}$, where $\rho_B$ is the baryon density,
 the above equation reads
\begin{equation}
{\frac{\partial}{\partial t}}(\rho_{B}\gamma)+\vec{\nabla} \cdot (\rho_{B}\gamma \vec{v})=0
\label{con}
\end{equation}
or
\begin{equation}
{\frac{\partial \rho_{B}}{\partial t}}+\gamma^{2}v \rho_{B}\Bigg({\frac{\partial v}
{\partial t}}+ \vec{v}\cdot \vec{\nabla} v\Bigg)+\vec{\nabla} \cdot (\rho_{B}\vec{v})=0
\label{newcon}
\end{equation}
The enthalpy per nucleon is given by \cite{land}:
\begin{equation}
dh=Tds+Vdp 
\label{ent1}
\end{equation}
where \hspace{0.2cm} $V=1/\rho_{B}$ \hspace{0.2cm} is the specific volume. $T$ and $s$ 
are the temperature and entropy density respectively.  
For a perfect fluid  $(ds=0)$    the  equation above becomes  $dp=\rho_{B}dh$ and 
consequently:
\begin{equation}
\vec{\nabla} p=\rho_{B}\vec{\nabla} h, \hspace{1cm} 
{\frac{\partial p}{\partial t}}=\rho_{B}{\frac{\partial h}{\partial t}}
\label{deripe}
\end{equation}
The Gibbs relation is \cite{reif}:
\begin{equation}
\varepsilon+p={\mu_{B}} {\rho_{B}}+Ts
\label{070}
\end{equation} 
where  $\mu_{B}$ is the baryochemical potential.  Inserting (\ref{deripe}) and 
(\ref{070}) in (\ref{eul}) we find:
\begin{equation}
{\frac{\partial {\vec{v}}}{\partial t}}+(\vec{v} \cdot \vec{\nabla})\vec{v}=
-{\frac{\rho_{B}}{({\mu_{B}} {\rho_{B}}+Ts)\gamma^{2}}}
\bigg(\vec{\nabla} h  +\vec{v} {\frac{\partial h}{\partial t}}\bigg)
\label{neweuler}
\end{equation}
The enthalpy per nucleon can also be calculated with the expression \cite{abu}:
\begin{equation}
h=E+\rho_{B}{\frac{\partial E}{\partial\rho_{B}}} 
\label{hcomE}
\end{equation} 
where $E$ is the energy per nucleon given by:
$$
E=\frac{\varepsilon}{\rho_{B}}
$$
which, inserted into  ({\ref{hcomE}}) yields:
\begin{equation}
h={\frac{\partial \varepsilon }{\partial \rho_{B}}} 
\label{068}
\end{equation}
It is  clear that the ``force'' on the 
right hand side of  ({\ref{neweuler}})  will be ultimately  
determined by the equation of state, i.e., by the function $\varepsilon(\rho_B)$. 

\section{Equation of state}

Equation  ({\ref{neweuler}})   contains the gradient of the 
derivative of the energy density. If $\varepsilon$ contains a 
Laplacian of $\rho_B$, i.e., ${\mathcal{\varepsilon}} 
\propto ... + ... \nabla^{2} \rho_{B} + ...$, then 
({\ref{neweuler}})  will have a cubic derivative
with respect to the space coordinate, which will give rise to the Korteweg-de Vries equation 
for the baryon density. 
The most popular  relativistic mean field models do not have higher derivative terms and, 
even if they have at the start, these terms are usually neglected during the calculations.  

As in \cite{fn1} we shall use a variant of  the non-linear Walecka model \cite{lala}  
given by:
\begin{equation}
\mathcal{L}=\mathcal{L}_{QHD}+{\frac{ d \,g_{V}}{{m_{V}}^{2}}}\bar{\psi}
(\partial_{\nu} \partial^{\nu} V_{\mu})\gamma^{\mu} \psi 
\label{lagr}
\end{equation}
with
$$
\mathcal{L}_{QHD}=\bar{\psi}[\gamma_{\mu}(i \partial^{\mu} - g_{V}V^{\mu})-(M-g_{S} \phi)]\psi + 
{\frac{1}{2}}\Big(\partial_{\mu} \phi \partial^{\mu} \phi - {m_{S}}^{2} \phi^{2}\Big)+
$$
$$
-{\frac{b{\phi}^{3}}{3}}-{\frac{c{\phi}^{4}}{4}} 
-{\frac{1}{4}}F_{\mu \nu}F^{\mu \nu} +{\frac{1}{2}}{m_{V}}^{2}V_{\mu}V^{\mu}
$$
where $F_{\mu \nu} =  \partial_{\mu} V_{\nu} - \partial_{\nu} V_{\mu}$. 
As usual, the degrees of freedom are 
the baryon field $\psi$, the neutral scalar meson field $\phi$
and the neutral vector meson field $V_{\mu}$, with the respective couplings and masses. 
The second and new term in (\ref{lagr}) is designed to be small in comparison with the main 
baryon - vector meson interaction term $g_{v} \bar{\psi} \gamma_{\mu} V^{\mu}  \psi$. 
Because of the derivatives, it is of the 
order of:
\begin{equation}
\frac{p^2}{m_V^2} \sim \frac{k_F^2}{m_V^2} \sim 0.12
\label{estimate}
\end{equation}
where the Fermi momentum is $k_F\simeq 0.28$ GeV and $m_V \simeq 0.8$ GeV. 
The form chosen for the new interaction term is not dictated by any symmetry argument, has no 
other deep justification and is just one possible interaction term among many others.  It 
is used here as a  prototype. At this stage our main interest is to explore the effects of 
these higher derivative terms, which may generate more complex wave equations. The parameter 
$d$ is free and will act as a ``marker''. Setting $d$ equal to zero will 
switch off the new term. On the other hand $d=1$ means that the coupling $g_V$ is the standard 
one. Other values imply a correction in this coupling. 

The mean field approximation means that  
$V_{\mu} \rightarrow <V_{\mu}> \equiv \delta_{\mu 0} V_{0}$  and 
$\phi \rightarrow <\phi> \equiv \phi_{0} $.  
With the above Lagrangian and the corresponding Hamiltonian we can, following the 
standard procedure,  write the partition 
function of the system and calculate the  energy density, the pressure and entropy density,  
which, for symmetric nuclear matter,  are  given by \cite{amigos}:
$$
\varepsilon={\frac{b}{3{g_{S}}^{3}}}(M-M^{*})^{3} 
+{\frac{{c}}{4{g_{S}}^{4}}}
(M-M^{*})^{4}+{\frac{\gamma_s}{(2\pi)^{3}}}\int d^3{k}\hspace{0.2cm}h_{+}\hspace{0.2cm}[n_{\vec{k}}(T,\nu)
+\bar{n}_{\vec{k}}(T,\nu)]
$$
\begin{equation}
+{\frac{{g_{V}}^{2}}{2{m_{V}}^{2}}}{\rho_{B}}^{2}+
{\frac{{m_{S}}^{2}}{2{g_{S}}^{2}}}(M-M^{*})^{2}
+{\frac{{ d \, g_{V}}^{2}}{{m_{V}}^{4}}}\rho_{B}\vec{\nabla}^{2}\rho_{B} 
-{\frac{{d \, g_{V}}^{2}}{{m_{V}}^{4}}}\rho_{B}
\bigg({\frac{\partial^{2}\rho_{B}}{\partial t^{2}}}\bigg)
\label{epst}
\end{equation}
$$
p={\frac{{g_{V}}^{2}}{2{m_{V}}^{2}}}{\rho_{B}}^{2}-{\frac{{m_{S}}^{2}}{2{g_{S}}^{2}}}(M-M^{*})^{2}
-{\frac{b}{3{g_{S}}^{3}}}(M-M^{*})^{3} 
-{\frac{c}{4{g_{S}}^{4}}}(M-M^{*})^{4}
$$
$$
-T{\frac{\gamma_s}{(2\pi)^{3}}}\int d^3{k}\hspace{0.2cm}\bigg\lbrace ln[(1-n_{\vec{k}}(T,\nu))]+
ln[(1-\bar{n}_{\vec{k}}(T,\nu))] \bigg\rbrace
$$
\begin{equation}
-{\frac{{d \, g_{V}}^{2}}{{m_{V}}^{4}}}\rho_{B}\vec{\nabla}^{2}\rho_{B}
+{\frac{{d \, g_{V}}^{2}}{{m_{V}}^{4}}}\rho_{B}
\bigg({\frac{\partial^{2}\rho_{B}}{\partial t^{2}}}\bigg)
\label{press}
\end{equation}  

and
$$
s=-{\frac{\gamma_s}{(2\pi)^{3}}}\int d^3{k}\bigg\lbrace n_{\vec{k}}(T,\nu)ln\Big[n_{\vec{k}}(T,\nu)\Big]+
\Big[1-n_{\vec{k}}(T,\nu)\Big]ln\Big[1-n_{\vec{k}}(T,\nu)\Big]+
$$
\begin{equation}
+\bar{n}_{\vec{k}}(T,\nu)
ln\Big[\bar{n}_{\vec{k}}(T,\nu)\Big]+\Big[1-\bar{n}_{\vec{k}}(T,\nu)\Big]ln\Big[1-\bar{n}_{\vec{k}}(T,\nu)\Big]\bigg\rbrace
\label{037}
\end{equation}
where $\gamma_s=4$ is the degeneracy factor and  $M^*$ is the effective nucleon mass 
( $M^{*}= M - g_{S} \phi_{0}$) given by: 
$$
M^{*}=M-{\frac{{g_{S}}^{2}}{{m_{S}}^{2}}}{\frac{\gamma_s}{(2\pi)^{3}}}\int d^3{k}
{\frac{M^{*}}{h_{+}}}
[n_{\vec{k}}(T,\nu)
+\bar{n}_{\vec{k}}(T,\nu)]+
$$
\begin{equation}
+{\frac{{g_{S}}^{2}}{{m_{S}}^{2}}}\bigg[{\frac{b}{{g_{S}}^{3}}}(M-M^{*})^{2}+{\frac{c}{{g_{S}}^{4}}}(M-M^{*})^{3}\bigg]
\label{045}
\end{equation}
and 
\begin{equation}
\rho_{B}={\frac{\gamma_s}{(2\pi)^{3}}}\int d^3{k}\hspace{0.2cm}
[n_{\vec{k}}(T,\nu)-\bar{n}_{\vec{k}}(T,\nu)]
\label{032}
\end{equation}
with
\begin{equation}
n_{\vec{k}}(T,\nu)\equiv{\frac{1}{1+e^{(h_{+}-\nu)/T}}}
\label{031}
\end{equation}
\begin{equation}
\bar{n}_{\vec{k}}(T,\nu)\equiv{\frac{1}{1+e^{(h_{+}+\nu)/T}}}
\label{031aaa}
\end{equation}
\begin{equation}
\nu \equiv \mu_{B}-g_{V}V_{0}
\label{026}
\end{equation}
\begin{equation}
h_{+}\equiv({\vec{k}}^{2}+{M^{*}}^{2})^{1/2}
\label{019}
\end{equation}
The last  two terms of (\ref{epst}) come from  the new interaction term in (\ref{lagr}). 
Baryon number propagation in nuclear matter has been studied in \cite{shin} with the help 
of the diffusion equation: 
\begin{equation}
{\frac{\partial \rho_{B}}{\partial t}} = D \, \nabla^{2}\rho_{B}
\label{diffusion}
\end{equation}
where the diffusion constant $D$ was numerically evaluated as a function of density and 
temperature and found to be $D\simeq 0.35 fm$ at densities comparable to the equilibrium 
nuclear density and temperatures of the order of $80$ MeV.  This number 
is small compared to any nuclear size scale and can be interpreted  as indicating that 
density gradients do not disappear very rapidly in nucler matter.
Using (\ref{diffusion}) twice in the
last term of (\ref{epst}) it can be rewritten as:
\begin{equation}
-{\frac{{d \, g_{V}}^{2}}{{m_{V}}^{4}}}\rho_{B}
\bigg({\frac{\partial^{2}\rho_{B}}{\partial t^{2}}}\bigg)=
-{\frac{{d \, g_{V}}^{2}}{{m_{V}}^{4}}}\rho_{B}{\frac{\partial}{\partial t}}\bigg(D 
\nabla^{2}\rho_{B}\bigg) = 
-{\frac{{d \, g_{V}}^{2}}{{m_{V}}^{4}}}\rho_{B}D^{2}[\nabla^{2}(\nabla^{2}\rho_{B})]
\label{sectimederi}
\end{equation}
which, in the context of the present calculation, can be neglected  because
$\nabla^{2}(\nabla^{2}\rho_{B})<<(\nabla^{2}\rho_{B})$. 
With this last approximation, the final form of the energy density is given by 
(\ref{epst}) without the last term. Of course, the same argument holds for the pressure, 
which will be  given by (\ref{press}) without the last term.

When  $T=0$ the energy density reduces to:
$$
\varepsilon={\frac{{g_{V}}^{2}}{2{m_{V}}^{2}}}{\rho_{B}}^{2}
+{\frac{{m_{S}}^{2}}{2{g_{S}}^{2}}}(M-M^{*})^{2}
+{\frac{{d \, g_{V}}^{2}}{{m_{V}}^{4}}}\rho_{B}\vec{\nabla}^{2}\rho_{B}+
{\frac{b}{3{g_{S}}^{3}}}(M-M^{*})^{3}+
$$
\begin{equation}
+{\frac{{c}}{4{g_{S}}^{4}}}
(M-M^{*})^{4}+{\frac{\gamma_s}{(2\pi)^{3}}}\int_{0}^{k_F} d^3{k}\hspace{0.2cm}h_{+}\hspace{0.2cm}
\label{epstzero}
\end{equation}
From ({\ref{epst}}) + ({\ref{press}}) we can  check that the Gibbs relation (\ref{070}) 
is fulfilled. The speed of sound $c_{s}$ is given by:
\begin{equation}
{c_{s}}^{2}={\frac{\partial p}{\partial \varepsilon}}
\label{054}
\end{equation}

The numerical inputs for the above formulas for $T=0$ were  taken from \cite{lala} and are shown 
in  Table I. In the Table the incompressibility $K$, the effective nucleon mass $M^*$, the 
speed of sound $c_s$ and the saturation density $\rho_0$ are calculated.  For $T > 0$ 
the sign of the parameter $c$ is reversed. In order to test our routines we have reproduced 
the results shown in Table I, but in what follows all the results will be obtained with the 
parameter set NL1.

\begin{center}
\begin{tabular}{|l|l|l|l|l|l|r|}  \hline  \hline  
         & $QHD$ & $NL1$ & $NL3$   & $NL3-II$ & $NL-SH$       \\ \hline  
K(MeV)   & 545       & 211   & 272  & 272  & 355        \\ \hline
M(MeV)   & 939       & 938   & 939     & 939      & 939           \\ \hline
$m_{S}(MeV)$   & 500   & 492  & 508,2 & 507,7 & 526    \\ \hline
$m_{V}(MeV)$   & 780   & 783 & 782,5 & 781,9 & 783            \\ \hline
$g_{S}$   & 8,7   & 10,14  & 10,22 & 10,2 & 10,4   \\ \hline
$g_{V}$   & 11,62  & 13,28 & 12,87 & 12,8 & 12,9    \\ \hline
$M^{*}/{M}$ & 0,56   & 0,57  & 0,6 & 0,59 & 0,6   \\ \hline
${\rho_{0}}(fm^{-3})$ & 0,19   & 0,15  & 0,15 & 0,15 & 0,15   \\ \hline
$b(fm^{-1})$ & 0   & +12,17  & +10,43 & +10,4 & +6,9  \\ \hline
$c$  & 0 & -36,26 & -28,9 & -28,9 & -15,8  \\ \hline
$c_{s}$ & 0,25   & 0,16 & 0,18 & 0,18 & 0,2   \\ \hline \hline  
\end{tabular}
\end{center}
\begin{center}
\small {Table I: Numerical inputs \cite{lala} for the equation of state.  $K$,  $M^*$, 
 $\rho_0$ and $c_{s}$ are calculated.} 
\end{center}

\section{Derivation of the Spherical KdV equation}

\subsection{Zero temperature}

This section contains  the derivation of the spherical KdV equation. The general 
scheme is the same as the one used in \cite{fn1} for the one dimensional Cartesian 
problem. However, as it wil be seen,  there are new details, which deserve the discussion 
presented below. We shall be concerned only with problems which have spherical symmetry. 
Therefore the continuity equation (\ref{newcon}) and Euler equation ({\ref{neweuler}) will 
have only radial components and the derivatives with respect to angles will vanish. 

Cold nuclear matter exhibits the saturation property, i.e., the energy per nucleon 
as a function of the baryon density ($E=\varepsilon/\rho_{B}$) has a minimum.  
Thus we  start with (\ref{epstzero}) and impose the saturation condition:
\begin{equation}
{\frac{\partial }{{\partial\rho_{B}}}}\bigg({\frac{\varepsilon}
{\rho_{B}}}-M\bigg)_{\rho_{B}=\rho_{0}}\hspace{0.5cm}=\hspace{0.5cm}0
\label{satcondagain}
\end{equation}
We then perform a Taylor expansion of $E$ around the equilibrium density $\rho_0$
up to second order: 
\begin{equation}
E(\rho_{B})=E(\rho_{0})
+\frac{1}{2}  
\bigg({\frac{\partial^{2} E}{{\partial\rho_{B}}^{2}}}\bigg)_{\rho_{B}=\rho_{0}} 
(\rho_{B}-\rho_{0})^{2}   
\label{expandagain}
\end{equation}
As in \cite{abu}, the density $\rho_B$ and its gradient $ \vec{\nabla} \rho_{B}$ 
are treated as independent variables. Inserting the above expression into (\ref{hcomE}) 
and using the relation \cite{abu}
\begin{equation}
\bigg({\frac{\partial^{2} E}{{\partial\rho_{B}}^{2}}}\bigg)_{\rho_{B}=\rho_{0}}  = 
\frac{M{c_{s}}^2}{\rho_{0}^2} 
\label{relation}
\end{equation} 
we obtain:
$$
h={\frac{{g_{V}}^{2}}{2{m_{V}}^{2}}}\rho_{0} 
+{\frac{{m_{S}}^{2}}{2\rho_{0}}}{\bigg[{\frac{(M^{*}-M)}{g_{S}}}\bigg]}^{2}
+{\frac{\gamma_s}{(2\pi)^{3}\rho_{0}}}\int_{0}^{k_{F}} d^3{k} ({\vec{k}}^{2}+{M^{*}}^{2})^{1/2}+
$$
$$
+{\frac{b}{3{g_{S}}^{3}\rho_{0}}}(M-M^{*})^{3}+{\frac{{c}}{4{g_{S}}^{4}\rho_{0}}}(M-M^{*})^{4}+
d{\frac{{g_{V}}^{2}}{{m_{V}}^{4}}}\vec{\nabla}^{2}\rho_{B}+
$$
\begin{equation}
+{\frac{M{c_{s}}^{2}}{2{\rho_{0}}^{2}}}(3{\rho_{B}}^{2}-4\rho_{B}\rho_{0}+{\rho_{0}}^{2})
\label{agah}
\end{equation} 
With the above expression we compute the derivatives: 
$$
{\frac{\partial h}{\partial r}}=-{\frac{2M{c_{s}}^{2}}{\rho_{0}}}{\frac{\partial\rho_{B}}
{\partial r}}
+{\frac{3M{c_{s}}^{2}}{{\rho_{0}}^{2}}}{\rho_{B}}{\frac{\partial\rho_{B}}{\partial r}}+
d{\frac{{g_{V}}^{2}}{{m_{V}}^{4}}}{\frac{\partial^{3} \rho_{B}}{\partial r^{3}}}+
$$
\begin{equation}
+d{\frac{{g_{V}}^{2}}{{m_{V}}^{4}}}{\frac{2}{r}}
{\frac{\partial^{2} \rho_{B}}{\partial r^{2}}} 
-d{\frac{{g_{V}}^{2}}{{m_{V}}^{4}}}{\frac{2}{r^{2}}}
{\frac{\partial \rho_{B}}{\partial r}}    
\label{gradradial}
\end{equation} 
and also:
$$
{\frac{\partial h}{\partial t}}=-{\frac{2M{c_{s}}^{2}}{\rho_{0}}}{\frac{\partial\rho_{B}}{\partial t}}
+{\frac{3M{c_{s}}^{2}}{{\rho_{0}}^{2}}}{\rho_{B}}{\frac{\partial\rho_{B}}{\partial t}}
+d{\frac{{g_{V}}^{2}}{{m_{V}}^{4}}}
{\frac{\partial}{\partial t}}\bigg({\frac{\partial^{2} \rho_{B}}{\partial r^{2}}}\bigg)+
$$
\begin{equation}
+d{\frac{{g_{V}}^{2}}{{m_{V}}^{4}}}{\frac{2}{r}}{\frac{\partial}{\partial t}}
\bigg({\frac{\partial \rho_{B}}{\partial r}}\bigg) 
-d{\frac{{g_{V}}^{2}}{{m_{V}}^{4}}}{\frac{2v}{r^{2}}}{\frac{\partial \rho_{B}}{\partial r}}    
\label{tradial}
\end{equation}
Inserting  (\ref{gradradial}) and (\ref{tradial}) into (\ref{neweuler}) we find:
$$
{\frac{\partial {v}}{\partial t}}+v{\frac{\partial {v}}{\partial r}}=
{\frac{(v^{2}-1)}{\mu_{B}}}
\bigg\lbrace\bigg({\frac{3M{c_{s}}^{2}}{{\rho_{0}}^{2}}} {\rho_{B}}-{\frac{2M{c_{s}}^{2}}
{\rho_{0}}}\bigg)\bigg({\frac{\partial\rho_{B}}{\partial r}}
+v{\frac{\partial\rho_{B}}{\partial t}}\bigg)+
d{\frac{{g_{V}}^{2}}{{m_{V}}^{4}}}\bigg[{\frac{\partial^{3} \rho_{B}}{\partial r^{3}}}+
v{\frac{\partial}{\partial t}}\bigg({\frac{\partial^{2} \rho_{B}}
{\partial r^{2}}}\bigg)\bigg]+
$$
\begin{equation}
+d{\frac{{2g_{V}}^{2}}{r{m_{V}}^{4}}}\bigg[{\frac{\partial^{2} \rho_{B}}{\partial r^{2}}}-{\frac{1}{r}}
{\frac{\partial\rho_{B}}{\partial r}}+v{\frac{\partial}{\partial t}}
\bigg({\frac{\partial\rho_{B}}{\partial r}} \bigg)-{\frac{v^{2}}{r}}
{\frac{\partial\rho_{B}}{\partial r}}\bigg]\bigg\rbrace
\label{eulerenddimumradial}
\end{equation}
We now rewrite (\ref{eulerenddimumradial}) e (\ref{newcon}) in terms of the 
dimensionless variables:
\begin{equation}
\hat\rho={\frac{\rho_{B}}{\rho_{0}}} \hspace{0.2cm}, \hspace{0.5cm} \hat v={\frac{v}{c_{s}}}
\label{varschapeuagain}
\end{equation}
where  $\rho_{0}$ is the equilibrium baryon density given in Table I and  
$c_{s}$ is the speed of  
sound. The Euler equation becomes:
$$
c_{s}{\frac{\partial {\hat v}}{\partial t}}+{c_{s}}^{2}\hat v
{\frac{\partial {\hat v}}{\partial r}}=
{\frac{({c_{s}}^{2}\hat v^{2}-1)}{\mu_{B}}}
\bigg\lbrace\bigg({\frac{3M{c_{s}}^{2}}{{\rho_{0}}^{2}}} \rho_{0}{\hat\rho}-{\frac{2M{c_{s}}^{2}}{\rho_{0}}}\bigg)\rho_{0}\bigg({\frac{\partial\hat\rho}{\partial r}}
+c_{s}\hat v{\frac{\partial\hat\rho}{\partial t}}\bigg)+
d{\frac{{g_{V}}^{2}}{{m_{V}}^{4}}}\rho_{0}\bigg[{\frac{\partial^{3} \hat\rho}{\partial r^{3}}}+
$$
\begin{equation}
c_{s} \hat v{\frac{\partial}{\partial t}}\bigg({\frac{\partial^{2} \hat\rho}
{\partial r^{2}}}\bigg)\bigg]
+d{\frac{2{g_{V}}^{2}}{r{m_{V}}^{4}}}\rho_{0}\bigg[{\frac{\partial^{2} \hat\rho}{\partial r^{2}}}-{\frac{1}{r}}
{\frac{\partial\hat\rho}{\partial r}}+c_{s}\hat v{\frac{\partial}{\partial t}}
\bigg({\frac{\partial\hat\rho}{\partial r}} \bigg)-{\frac{{c_{s}}^{2}\hat v^{2}}{r}}
{\frac{\partial\hat\rho}{\partial r}}\bigg]\bigg\rbrace
\label{euleradmradial}
\end{equation}
and the continuity equation becomes:
\begin{equation}
(1-{c_{s}}^{2}{\hat v}^{2})\bigg({\frac{\partial \hat\rho}{\partial t}}
+c_{s}\hat\rho{\frac{\partial \hat v}
{\partial r}}+c_{s}\hat v {\frac{\partial \hat\rho}{\partial r}}+
{\frac{2c_{s}\hat\rho  \hat v}{r}}\bigg)+
{c_{s}}^{2}\hat\rho  \hat v
\bigg({\frac{\partial \hat v}{\partial t}}+
c_{s}\hat v{\frac{\partial \hat v}{\partial r}}\bigg)=0
\label{contiadmradial}
\end{equation}
We next  define the ``stretched coordinates''  $\xi$ and $\tau$ as in 
\cite{frsw,rww,abu,davidson}:
\begin{equation}
\xi=\sigma^{1/2}{\frac{(r-{c_{s}}t)}{R}} 
\hspace{0.2cm}, \hspace{0.5cm} 
\tau=\sigma^{3/2}{\frac{{c_{s}}t}{R}} 
\label{stretradial}       
\end{equation}
where $R$ is a size scale and $\sigma$ is a small ($0 < \sigma < 1$) expansion parameter.  
The derivatives become the following operators:
$$
{\frac{\partial}{\partial r}}={\frac{\sigma^{1/2}}{R}}{\frac{\partial}{\partial \xi}} 
\hspace{0.3cm}, \hspace{0.5cm} 
{\frac{\partial^{2}}{\partial r^{2}}}={\frac{\sigma}{R^{2}}}
{\frac{\partial^{2}}{\partial \xi^{2}}} 
\hspace{0.3cm}, \hspace{0.5cm}
{\frac{\partial^{3}}{\partial r^{3}}}={\frac{\sigma^{3/2}}{R^{3}}}
{\frac{\partial^{3}}{\partial \xi^{3}}} 
\hspace{0.4cm} and 
$$
\begin{equation}
{\frac{\partial}{\partial t}}=
-{\frac{\sigma^{1/2}{c_{s}}}{R}}{\frac{\partial}{\partial \xi}}
+{\frac{\sigma^{3/2}{c_{s}}}{R}}{\frac{\partial}{\partial \tau}}
\label{stretopersradiais}       
\end{equation}
From (\ref{stretradial}) we can see that:
\begin{equation}
r={\frac{R \xi \sigma + R \tau}{\sigma^{3/2}}}
\label{rstret}
\end{equation}
and thus:
\begin{equation}
{\frac{1}{r^{2}}}={\frac{\sigma^{3}}{(R \xi \sigma + R \tau)^{2}}}
\label{umrquad}
\end{equation}
In the $\xi-\tau$ space   (\ref{euleradmradial}) reads: 
$$
-{\frac{\sigma^{1/2}{c_{s}}^{2}}{R}}{\frac{\partial \hat v}{\partial \xi}}
+{\frac{\sigma^{3/2}{c_{s}}^{2}}{R}}{\frac{\partial \hat v}{\partial \tau}}+
{c_{s}}^{2}{\frac{\sigma^{1/2}}{R}}\hat v{\frac{\partial \hat v}{\partial \xi}}=
$$
$$
={\frac{({c_{s}}^{2}{\hat v}^{2}-1)}{\mu_{B}}}
\bigg\lbrace\bigg({\frac{3M{c_{s}}^{2}}{{\rho_{0}}^{2}}} {\rho_{0}}\hat\rho-{\frac{2M{c_{s}}^{2}}{\rho_{0}}}\bigg)\rho_{0}
\bigg({\frac{\sigma^{1/2}}{R}}{\frac{\partial \hat \rho}{\partial \xi}}
-{\frac{\sigma^{1/2}{c_{s}}^{2}}{R}}\hat v{\frac{\partial\hat\rho}{\partial \xi}}
+{\frac{\sigma^{3/2}{c_{s}}^{2}}{R}}\hat v{\frac{\partial\hat\rho}{\partial \tau}}\bigg)+
$$
$$
+d{\frac{{g_{V}}^{2}}{{m_{V}}^{4}}} \rho_{0}\bigg[{\frac{\sigma^{3/2}}{R^{3}}}
{\frac{\partial^{3}\hat\rho}{\partial \xi^{3}}} +
\bigg(-{\frac{\sigma^{1/2}{c_{s}}^{2}}{R}}\hat v{\frac{\partial}{\partial \xi}}
+{\frac{\sigma^{3/2}{c_{s}}^{2}}{R}}\hat v{\frac{\partial}{\partial \tau}}\bigg)
\bigg({\frac{\sigma}{R^{2}}}
{\frac{\partial^{2}\hat\rho}{\partial \xi^{2}}}\bigg)\bigg]+
$$
$$
+d{\frac{2{g_{V}}^{2}\rho_{0}\sigma^{3/2}}{{{m_{V}}^{4}(R \xi \sigma + R \tau)}}}\bigg[
{\frac{\sigma}{R^{2}}}{\frac{\partial^{2} \hat\rho}{\partial \xi^{2}}}-
{\frac{\sigma^{3/2}}{(R \xi \sigma + R \tau)}}{\frac{\sigma^{1/2}}{R}}
{\frac{\partial\hat\rho}{\partial \xi}}+
$$
\begin{equation}
+c_{s}\hat v
\bigg(-{\frac{\sigma^{1/2}{c_{s}}}{R}}{\frac{\partial}{\partial \xi}}
+{\frac{\sigma^{3/2}{c_{s}}}{R}}{\frac{\partial}{\partial \tau}}\bigg)
\bigg({\frac{\sigma^{1/2}}{R}}
{\frac{\partial\hat\rho}{\partial \xi}} \bigg)-{\frac{{c_{s}}^{2}\hat v^{2}\sigma^{3/2}}
{(R \xi \sigma + R \tau)}}{\frac{\sigma^{1/2}}{R}}
{\frac{\partial\hat\rho}{\partial \xi}}\bigg]\bigg\rbrace
\label{euleradmxitauradial}
\end{equation}
and the continuity equation (\ref{contiadmradial}) reads: 
$$
(1-{c_{s}}^{2}{\hat v}^{2})\bigg(
-{\frac{\sigma^{1/2}{c_{s}}}{R}}{\frac{\partial \hat\rho}{\partial \xi}}
+{\frac{\sigma^{3/2}{c_{s}}}{R}}{\frac{\partial \hat\rho}{\partial \tau}}
+c_{s}\hat\rho
{\frac{\sigma^{1/2}}{R}}{\frac{\partial \hat v}{\partial \xi}} 
+c_{s}\hat v {\frac{\sigma^{1/2}}{R}}{\frac{\partial \hat \rho}{\partial \xi}} 
+{\frac{2c_{s}\hat v \hat \rho \sigma^{3/2}}{(R \xi \sigma+R \tau)}}\bigg)+
$$
\begin{equation}
+{c_{s}}^{2}\hat\rho  \hat v
\bigg(-{\frac{\sigma^{1/2}{c_{s}}}{R}}{\frac{\partial \hat v}{\partial \xi}}
+{\frac{\sigma^{3/2}{c_{s}}}{R}}{\frac{\partial \hat v}{\partial \tau}}+
c_{s}\hat v{\frac{\sigma^{1/2}}{R}}{\frac{\partial \hat v}{\partial \xi}} \bigg)=0
\label{contiadmxitauradial}
\end{equation}
We then expand (\ref{varschapeuagain})  around  the equilibrium values:
\begin{equation}
\hat\rho=1+\sigma \rho_{1}+ \sigma^{2} \rho_{2}+ \dots = 1\, + \, \hat\rho_1 \, + \,  
\hat\rho_2  + \,  \dots
\label{roexpagain}
\end{equation}
\begin{equation}
\hat v=\sigma v_{1}+ \sigma^{2} v_{2}+ \dots
\label{vexpagain}
\end{equation}
With this expansion (\ref{euleradmxitauradial}) becomes:
$$
-{\frac{\sigma^{1/2}{c_{s}}^{2}}{R}}
{\frac{\partial}{\partial \xi}}(\sigma v_{1}+ \sigma^{2} v_{2}+ \dots)
+{\frac{\sigma^{3/2}{c_{s}}^{2}}{R}}
{\frac{\partial}{\partial \tau}}(\sigma v_{1}+ \sigma^{2} v_{2}+ \dots)+
$$
$$
+{c_{s}}^{2}
{\frac{\sigma^{1/2}}{R}}(\sigma v_{1}+ \sigma^{2} v_{2}+ \dots)
{\frac{\partial}{\partial \xi}}(\sigma v_{1}+ \sigma^{2} v_{2}+ \dots)=
$$
$$
={\frac{[{c_{s}}^{2}{(\sigma v_{1}+ \sigma^{2} v_{2}+ \dots)}^{2}-1]}{\mu_{B}}} 
\hspace{0.3cm}\times
$$
$$
\hspace{0.3cm} \times
\bigg\lbrace\bigg[{\frac{3M{c_{s}}^{2}}{{\rho_{0}}^{2}}} {\rho_{0}}(1+\sigma \rho_{1}+ 
\sigma^{2} \rho_{2}+ \dots)-{\frac{2M{c_{s}}^{2}}{\rho_{0}}}\bigg]\rho_{0}
\bigg[{\frac{\sigma^{1/2}}{R}}
{\frac{\partial}{\partial \xi}}(1+\sigma \rho_{1}+ \sigma^{2} \rho_{2}+ \dots)+
$$
$$
-{\frac{\sigma^{1/2}{c_{s}}^{2}}{R}}(\sigma v_{1}+ \sigma^{2} v_{2}+ \dots)
{\frac{\partial}{\partial \xi}}(1+\sigma \rho_{1}+ \sigma^{2} \rho_{2}+ \dots)+
$$
$$
+{\frac{\sigma^{3/2}{c_{s}}^{2}}{R}}
(\sigma v_{1}+ \sigma^{2} v_{2}+ \dots)
{\frac{\partial}{\partial \tau}}(1+\sigma \rho_{1}+ \sigma^{2} \rho_{2}+ \dots)\bigg]+
$$
$$
+d{\frac{{g_{V}}^{2}}{{m_{V}}^{4}}}\rho_{0}\bigg[{\frac{\sigma^{3/2}}{R^{3}}}
{\frac{\partial^{3}}{\partial \xi^{3}}}(1+\sigma \rho_{1}+ \sigma^{2} \rho_{2}+ \dots)+
\bigg(-{\frac{\sigma^{1/2}{c_{s}}^{2}}{R}}(\sigma v_{1}+ \sigma^{2} v_{2}+ \dots)
{\frac{\partial}{\partial \xi}}+
$$
$$
+{\frac{\sigma^{3/2}{c_{s}}^{2}}{R}}(\sigma v_{1}+ \sigma^{2} v_{2}+ \dots)
{\frac{\partial}{\partial \tau}}\bigg)
\bigg({\frac{\sigma}{R^{2}}}
{\frac{\partial^{2}}
{\partial \xi^{2}}}(1+\sigma \rho_{1}+ \sigma^{2} \rho_{2}+ \dots)
\bigg)\bigg]+
$$
$$
+d{\frac{2{g_{V}}^{2}\rho_{0}\sigma^{3/2}}{{{m_{V}}^{4}(R \xi \sigma + R \tau)}}}\bigg[
{\frac{\sigma}{R^{2}}}{\frac{\partial^{2}}
{\partial \xi^{2}}}(1+\sigma \rho_{1}+ \sigma^{2} \rho_{2}+ \dots)+
$$
$$
-{\frac{\sigma^{3/2}}{(R \xi \sigma + R \tau)}}{\frac{\sigma^{1/2}}{R}}
{\frac{\partial}{\partial \xi}}(1+\sigma \rho_{1}+ \sigma^{2} \rho_{2}+ \dots)+
$$
$$
+c_{s}(\sigma v_{1}+ \sigma^{2} v_{2}+ \dots)
\bigg(-{\frac{\sigma^{1/2}{c_{s}}}{R}}{\frac{\partial}{\partial \xi}}
+{\frac{\sigma^{3/2}{c_{s}}}{R}}{\frac{\partial}{\partial \tau}}\bigg)
\bigg({\frac{\sigma^{1/2}}{R}}
{\frac{\partial}{\partial \xi}}(1+\sigma \rho_{1}+ \sigma^{2} \rho_{2}+ \dots) \bigg)
$$
\begin{equation}
-{\frac{{c_{s}}^{2}(\sigma v_{1}+ \sigma^{2} v_{2}+ \dots)^{2}\sigma^{3/2}}
{(R \xi \sigma + R \tau)}}{\frac{\sigma^{1/2}}{R}}
{\frac{\partial}{\partial \xi}}(1+\sigma \rho_{1}+ \sigma^{2} \rho_{2}+ \dots)
\bigg]\bigg\rbrace
\label{euleradmxitauradialexpand}
\end{equation}
and (\ref{contiadmxitauradial}) becomes:
$$
[1-{c_{s}}^{2}{(\sigma v_{1}+ \sigma^{2} v_{2}+ \dots)}^{2}]\bigg(
-{\frac{\sigma^{1/2}{c_{s}}}{R}}{\frac{\partial}{\partial \xi}}
(1+\sigma \rho_{1}+ \sigma^{2} \rho_{2}+ \dots)+
$$
$$
+{\frac{\sigma^{3/2}{c_{s}}}{R}}{\frac{\partial}{\partial \tau}}
(1+\sigma \rho_{1}+ \sigma^{2} \rho_{2}+ \dots)+
$$
$$
+c_{s}(1+\sigma \rho_{1}+ \sigma^{2} \rho_{2}+ \dots)
{\frac{\sigma^{1/2}}{R}}{\frac{\partial}{\partial \xi}} 
(\sigma v_{1}+ \sigma^{2} v_{2}+ \dots)+
$$
$$
+c_{s}(\sigma v_{1}+ \sigma^{2} v_{2}+ \dots)
 {\frac{\sigma^{1/2}}{R}}{\frac{\partial}{\partial \xi}}
(1+\sigma \rho_{1}+ \sigma^{2} \rho_{2}+ \dots)+
$$
$$
+{\frac{2c_{s}(\sigma v_{1}+ \sigma^{2} v_{2}+ \dots)
(1+\sigma \rho_{1}+ \sigma^{2} \rho_{2}+ \dots)
\sigma^{3/2}}{(R \xi \sigma+R \tau)}}\bigg)+
$$
$$
+{c_{s}}^{2}(1+\sigma \rho_{1}+ \sigma^{2} \rho_{2}+ \dots)
(\sigma v_{1}+ \sigma^{2} v_{2}+ \dots)
\bigg[-{\frac{\sigma^{1/2}{c_{s}}}{R}}{\frac{\partial}{\partial \xi}}
(\sigma v_{1}+ \sigma^{2} v_{2}+ \dots)+
$$
\begin{equation}
+{\frac{\sigma^{3/2}{c_{s}}}{R}}{\frac{\partial}{\partial \tau}}
(\sigma v_{1}+ \sigma^{2} v_{2}+ \dots)+
c_{s}(\sigma v_{1}+ \sigma^{2} v_{2}+ \dots)
{\frac{\sigma^{1/2}}{R}}{\frac{\partial}{\partial \xi}}
(\sigma v_{1}+ \sigma^{2} v_{2}+ \dots) \bigg]=0
\label{contradialexpand}
\end{equation}
Since $\sigma$ is small we go only up to second order.  Therefore 
(\ref{euleradmxitauradialexpand}) and  (\ref{contradialexpand}) turn 
into:
$$
\sigma \bigg( -{\frac{\partial v_{1}}{\partial \xi}}+
{\frac{({\frac{3M{c_{s}}^{2}}{{\rho_{0}}^{2}}} \rho_{0}-{\frac{2M{c_{s}}^{2}}{\rho_{0}}})\rho_{0}}{\mu_{B}\hspace{0.1cm}{c_{s}}^{2}}}
{\frac{\partial \rho_{1}}{\partial \xi}}\bigg)+\sigma^{2}
\bigg[-{\frac{\partial v_{2}}{\partial \xi}}+{\frac{\partial v_{1}}{\partial \tau}}+
v_{1}{\frac{\partial v_{1}}{\partial \xi}}+
{\frac{({\frac{3M{c_{s}}^{2}}{{\rho_{0}}^{2}}} \rho_{0}-{\frac{2M{c_{s}}^{2}}
{\rho_{0}}})\rho_{0}}{\mu_{B}\hspace{0.1cm}{c_{s}}^{2}}}
{\frac{\partial \rho_{2}}{\partial \xi}}+
$$
\begin{equation}
{\frac{{\frac{3M{c_{s}}^{2}}{{\rho_{0}}^{2}}} {\rho_{0}}^{2}}{\mu_{B}\hspace{0.1cm}{c_{s}}^{2}}}
\rho_{1}{\frac{\partial \rho_{1}}{\partial \xi}}-
{\frac{({\frac{3M{c_{s}}^{2}}{{\rho_{0}}^{2}}} \rho_{0}-{\frac{2M{c_{s}}^{2}}{\rho_{0}}})\rho_{0}}{\mu_{B}}}
v_{1}{\frac{\partial \rho_{1}}{\partial \xi}}+
\bigg({\frac{d{\frac{{g_{V}}^{2}}{{m_{V}}^{4}}} \rho_{0}}{\mu_{B}\hspace{0.1cm}{c_{s}}^{2}R^{2}}}\bigg)
{\frac{\partial^{3}\rho_{1}}{\partial \xi^{3}}}\bigg]=0
\label{sigmafateulerrad}
\end{equation}
and
$$
\sigma \bigg( -{\frac{\partial \rho_{1}}{\partial \xi}}+
{\frac{\partial v_{1}}{\partial \xi}}\bigg)+\sigma^{2}
\bigg[{\frac{\partial v_{2}}{\partial \xi}}+{\frac{\partial \rho_{1}}{\partial \tau}}
-{c_{s}}^{2}v_{1}{\frac{\partial v_{1}}{\partial \xi}}-
{\frac{\partial \rho_{2}}{\partial \xi}}+
$$
$$
+v_{1}{\frac{\partial \rho_{1}}{\partial \xi}}+
\rho_{1}{\frac{\partial v_{1}}{\partial \xi}}+
{\frac{2}{(\xi \sigma+\tau)}}v_{1}\bigg]=0
$$
In the last term of the above expression, since $0 < \sigma < 1$, we  shall 
assume that $\tau > \xi \sigma$ and   make the approximation
\begin{equation}
{\frac{2}{(\xi \sigma+\tau)}} \cong {\frac{2}{\tau}}
\label{approx}
\end{equation}
and hence: 
\begin{equation}
\sigma \bigg( -{\frac{\partial \rho_{1}}{\partial \xi}}+
{\frac{\partial v_{1}}{\partial \xi}}\bigg)+\sigma^{2}
\bigg({\frac{\partial v_{2}}{\partial \xi}}+{\frac{\partial \rho_{1}}{\partial \tau}}
-{c_{s}}^{2}v_{1}{\frac{\partial v_{1}}{\partial \xi}}-
{\frac{\partial \rho_{2}}{\partial \xi}}+
v_{1}{\frac{\partial \rho_{1}}{\partial \xi}}+
\rho_{1}{\frac{\partial v_{1}}{\partial \xi}}+
{\frac{2}{\tau}}v_{1}\bigg)=0
\label{sigmafatcontrad}
\end{equation}
Since the coefficients in the above series are 
independent of each other we get a set of equations.  
From the terms proportional to $\sigma$ in (\ref{sigmafateulerrad}) and 
(\ref{sigmafatcontrad}) we find:
\begin{equation}
\rho_{1}=v_{1}
\label{rovrelatum}
\end{equation}
and also
\begin{equation}
{\frac{({\frac{3M{c_{s}}^{2}}{{\rho_{0}}^{2}}} \rho_{0}
-{\frac{2M{c_{s}}^{2}}{\rho_{0}}})\rho_{0}}{\mu_{B}\hspace{0.1cm}{c_{s}}^{2}}}
=1 
\label{condrad}
\end{equation}
and therefore
\begin{equation}
\mu_{B}=M
\label{condrad2}
\end{equation}
In fact, in (\ref{rovrelatum}) we might have an integration constant. However, as 
it was shown in \cite{rww} for the one dimensional Cartesian case, this would not 
 change the results significantly. For our purposes it is enough to consider 
(\ref{rovrelatum}), keeping in mind that it is only a particular solution of the problem. 
From the terms proportional to $\sigma^{2}$ in (\ref{sigmafateulerrad}) and
(\ref{sigmafatcontrad}), with the help of (\ref{rovrelatum}) and (\ref{condrad}),
we find:
$$
-{\frac{\partial \rho_{1}}{\partial \tau}}-
\rho_{1}{\frac{\partial \rho_{1}}{\partial \xi}}-
3\rho_{1}{\frac{\partial \rho_{1}}{\partial \xi}}+
{c_{s}}^{2}\rho_{1}{\frac{\partial \rho_{1}}{\partial \xi}}-
\bigg({\frac{d{\frac{{g_{V}}^{2}}{{m_{V}}^{4}}} \rho_{0}}{\mu_{B}\hspace{0.1cm}
{c_{s}}^{2}R^{2}}}\bigg)
{\frac{\partial^{3}\rho_{1}}{\partial \xi^{3}}}=
$$
$$
={\frac{\partial \rho_{1}}{\partial \tau}}
-{c_{s}}^{2}\rho_{1}{\frac{\partial \rho_{1}}{\partial \xi}}+
\rho_{1}{\frac{\partial \rho_{1}}{\partial \xi}}+
\rho_{1}{\frac{\partial \rho_{1}}{\partial \xi}}
+{\frac{2}{\tau}}\rho_{1}
$$
which,  after a rearrangement of terms and change of variables back  to the  $r-t$ space,
becomes the ``spherical KdV'' equation:
\begin{equation}
{\frac{\partial {\hat{\rho}_{1}}}{\partial t}}+
{c_{s}}{\frac{\partial {\hat{\rho}_{1}}}{\partial r}}+
(3-{c_{s}}^{2}){c_{s}}
{\hat{\rho}_{1}}{\frac{\partial{\hat{\rho}_{1}}}{\partial r}}
+d\bigg({\frac{{g_{V}}^{2}\rho_{0}}{2M{{m_{V}}^{4}}{c_{s}}}}\bigg)
{\frac{\partial^{3}{\hat{\rho}_{1}}}{\partial r^{3}}}+{\frac{{\hat{\rho}_{1}}}{t}}=0  
\label{esftzero}
\end{equation}
for which a suitable initial condition may be:
\begin{equation}
{\hat{\rho}_{1}}(r,t_{0})={\frac{3(u-{c_{s}})}{{c_{s}}}}(3-{c_{s}}^{2})^{-1}sech^{2}\bigg[
{m_{V}}^{2}\sqrt{{\frac{(u-{c_{s}}){c_{s}}M}
{2{g_{V}}^{2}{\rho_{0}}}}}(r-ut_{0}) \bigg] 
\label{initesf}
\end{equation}
This Gaussian-looking  form is motivated by the analytical solution of the KdV equation in 
one dimensional Cartesian coordinates discussed in \cite{fn1} . Here, the numbers $u$,  
$t_0$,... are parameters without special meaning.  

\subsection{Finite temperature}

Apart  from the trivial replacement of (\ref{epstzero}) by 
(\ref{epst}) there is another change when we consider nuclear 
matter at finite temperature. We do not restrict ourselves 
to the case where nuclear matter is saturated. Instead,  we shall
consider the case where there is an equilibrated background 
with constant density and zero velocity, upon which perturbations 
propagate, but no saturation. The difference is that with saturation, 
a system is bound and more stable, whereas in the present case 
stability is not guaranteed and this system might expand or shrink. 
In such a situation, perturbations would propagate in an expanding 
medium and the reference density $\rho_0$ in (\ref{varschapeuagain}) 
might change with time. This is the scenario that we have in heavy ion 
collisions at RHIC, which we plan to address in the future. Here we 
consider the simpler case of constant $\rho_0$. 

Substituting (\ref{epst}) into  (\ref{068})  we obtain:
\begin{equation}
h={\frac{{g_{V}}^{2}}{{m_{V}}^{2}}}{\rho_{B}}
+d{\frac{{g_{V}}^{2}}{{m_{V}}^{4}}}\vec{\nabla}^{2}\rho_{B}
\label{069}
\end{equation}
Using the definition of the operators in spherical coordinates 
we arrive at:
\begin{equation}
{\frac{\partial h}{\partial r}}={\frac{{g_{V}}^{2}}{{m_{V}}^{2}}}
{\frac{\partial\rho_{B}}{\partial r}}+
d{\frac{{g_{V}}^{2}}{{m_{V}}^{4}}}{\frac{\partial^{3} \rho_{B}}{\partial r^{3}}}+
d{\frac{{g_{V}}^{2}}{{m_{V}}^{4}}}{\frac{2}{r}}{\frac{\partial^{2} \rho_{B}}{\partial r^{2}}} 
-d{\frac{{g_{V}}^{2}}{{m_{V}}^{4}}}{\frac{2}{r^{2}}}{\frac{\partial \rho_{B}}{\partial r}}    
\label{072}
\end{equation} 
and also
\begin{equation}
{\frac{\partial h}{\partial t}}={\frac{{g_{V}}^{2}}{{m_{V}}^{2}}}{\frac{\partial\rho_{B}}{\partial t}}+
d{\frac{{g_{V}}^{2}}{{m_{V}}^{4}}}{\frac{\partial}{\partial t}}\bigg({\frac{\partial^{2} \rho_{B}}{\partial r^{2}}}\bigg)
+d{\frac{{g_{V}}^{2}}{{m_{V}}^{4}}}{\frac{2}{r}}{\frac{\partial}{\partial t}}
\bigg({\frac{\partial \rho_{B}}{\partial r}}\bigg) 
-d{\frac{{g_{V}}^{2}}{{m_{V}}^{4}}}{\frac{2v}{r^{2}}}{\frac{\partial \rho_{B}}{\partial r}}    
\label{073}
\end{equation}
Substituting (\ref{072}) and (\ref{073}) into (\ref{neweuler}) and repeating the steps 
described in the last section, i.e., introducing  dimensionless variables, changing   
variables to the  $\xi-\tau$ space, expanding $\hat{\rho}$ and $\hat{v}$ and  collecting the 
terms proportional to $\sigma$ and to $\sigma^2$ we obtain the following relations  from the 
Euler equation:
\begin{equation}
\sigma\bigg[-\bigg(\mu_{B}+{\frac{Ts}{\rho_{0}}}\bigg){\frac{\partial {{v_{1}}}}
{\partial \xi}}+
{\frac{{g_{V}}^{2}}{{m_{V}}^{2}}}{\frac{ \rho_{0}}{{c_{s}}^{2}}}
{\frac{\partial {{\rho_{1}}}}{\partial \xi}} \bigg]=0
\label{074}
\end{equation} 
and
$$
\sigma^{2}\bigg[\bigg(\mu_{B}+{\frac{Ts}{\rho_{0}}}\bigg)\bigg(-
{\frac{\partial {{v_{2}}}}{\partial \xi}}+
{\frac{\partial {{v_{1}}}}{\partial \tau}}+v_{1}{\frac{\partial {{v_{1}}}}
{\partial \xi}}\bigg)+
{\frac{{g_{V}}^{2}}{{m_{V}}^{2}}}{\frac{\rho_{0}}{{c_{s}}^{2}}}\rho_{1}
{\frac{\partial {{\rho_{1}}}}{\partial \xi}}+
$$

\begin{equation}
+{\frac{{g_{V}}^{2}}{{m_{V}}^{2}}}{\frac{\rho_{0}}{{c_{s}}^{2}}}{\frac{\partial 
{{\rho_{2}}}}{\partial \xi}}
-\mu_{B}\rho_{1}{\frac{\partial {{v_{1}}}}{\partial \xi}}
-{\frac{{g_{V}}^{2}}{{m_{V}}^{2}}}\rho_{0}v_{1}{\frac{\partial {{\rho_{1}}}}{\partial \xi}}
+{\frac{{g_{V}}^{2}}{{m_{V}}^{4}}}{\frac{\rho_{0}}{{c_{s}}^{2}R^{2}}}{\frac{\partial^{3}
{{\rho_{1}}}}{\partial \xi^{3}}}\bigg]=0
\label{075}
\end{equation} 
After some manipulations, the continuity equation  (\ref{newcon}) is written as:
\begin{equation}
(1-v^{2})\bigg({\frac{\partial \rho_{B}}{\partial t}}+\rho_{B}{\frac{\partial v}
{\partial r}}+v {\frac{\partial \rho_{B}}{\partial r}}+
{\frac{2\rho_{B}v}{r}}\bigg)+v\rho_{B}
\bigg({\frac{\partial v}{\partial t}}+v{\frac{\partial v}{\partial r}}\bigg)=0
\label{076}
\end{equation}
which, after the change of variables and expansion yields the following relations:
\begin{equation}
\sigma \bigg( -{\frac{\partial \rho_{1}}{\partial \xi}}+
{\frac{\partial v_{1}}{\partial \xi}}\bigg)=0
\label{077}
\end{equation}
and
\begin{equation}
\sigma^{2}
\bigg({\frac{\partial v_{2}}{\partial \xi}}+{\frac{\partial \rho_{1}}{\partial \tau}}
-{c_{s}}^{2}v_{1}{\frac{\partial v_{1}}{\partial \xi}}-
{\frac{\partial \rho_{2}}{\partial \xi}}+
v_{1}{\frac{\partial \rho_{1}}{\partial \xi}}+
\rho_{1}{\frac{\partial v_{1}}{\partial \xi}}+
{\frac{2}{\tau}}v_{1}\bigg)=0
\label{sigmafatcontradagainn}
\end{equation}
where in the last term of the above expression we made again the 
approximation (\ref{approx}).
From (\ref{074}) and (\ref{077}) we get the relations:
$$
\bigg(\mu_{B}+{\frac{Ts}{\rho_{0}}}\bigg)=
{\frac{{g_{V}}^{2}}{{m_{V}}^{2}}}{\frac{\rho_{0}}{{c_{s}}^{2}}}
$$
and
$$
{v_{1}}={\rho_{1}}
$$
Substituting these expressions in (\ref{075}) and (\ref{sigmafatcontradagainn}) and
combining the resulting equations we arrive at the finite temperature spherical KdV 
equation:
\begin{equation}
{\frac{\partial {\hat{\rho_{1}}}}{\partial t}}+{c_{s}}{\frac{\partial {\hat{\rho_{1}}}}
{\partial r}}
+\bigg(2-{c_{s}}^{2}
-{\frac{{\mu_{B}}{m_{V}}^{2}{c_{s}}^{2}}{2{g_{V}}^{2}{\rho_{0}}}}\bigg){c_{s}}
\hat{\rho_{1}}{\frac{\partial{\hat{\rho_{1}}}}{\partial r}}
+d\bigg({\frac{{c_{s}}}{2{{m_{V}}^{2}}}}\bigg)
{\frac{\partial^{3}{\hat{\rho_{1}}}}{\partial r^{3}}}+{\frac{\hat{\rho_{1}}}{t}}=0
\label{esft}
\end{equation}
In the numerical studies of this equation we have used an initial condition with  the 
form given by (\ref{initesf}) with several choices for the parameters.

\begin{figure}[t]
\includegraphics[scale=0.70]{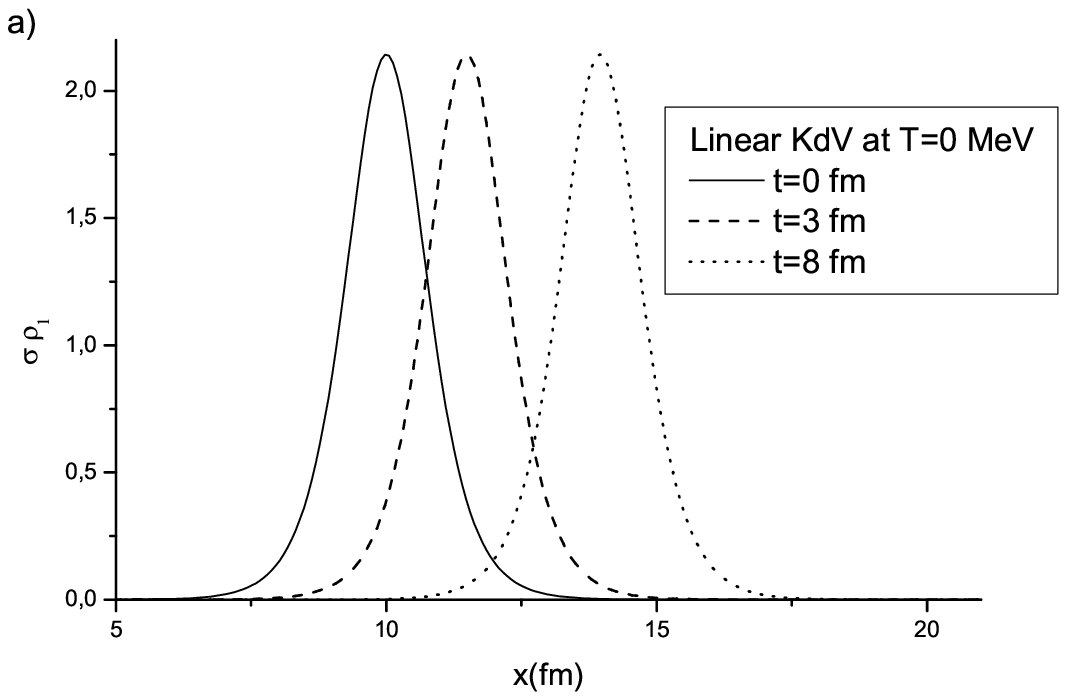}\\
\includegraphics[scale=0.70]{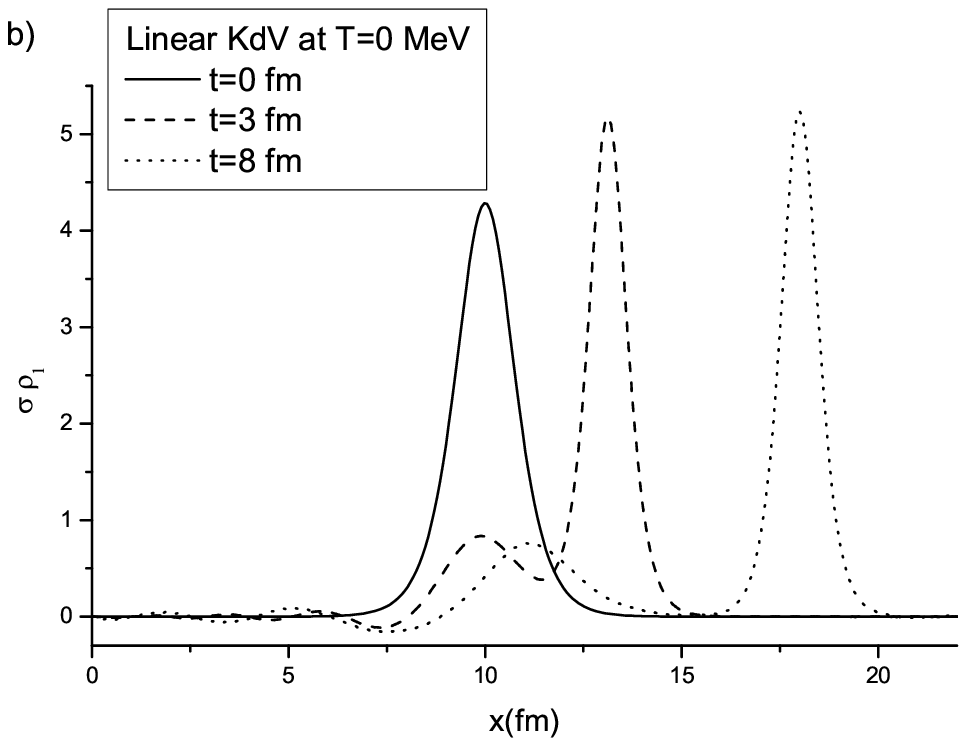}\\
\includegraphics[scale=0.70]{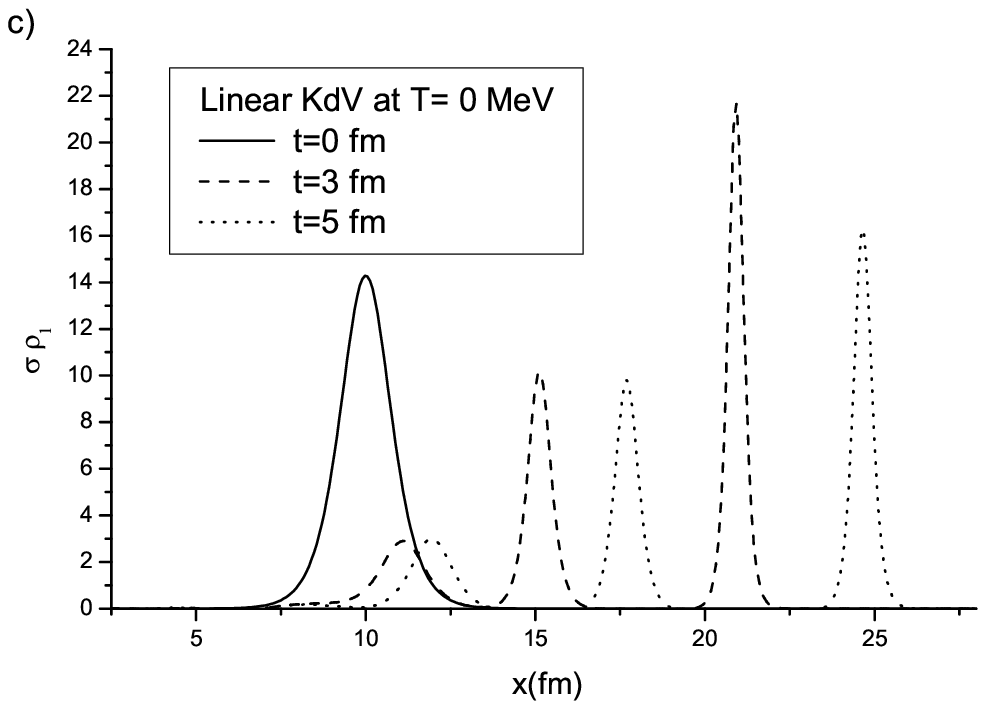}
\caption{Time evolution of a density perturbation in arbitrary units in one dimensional 
Cartesian coordinates. The upper pannel shows the evolution of the analytic solution of the 
KdV equation. In the lower pannels we show the evolution of the analytic solution
multiplied by a factor 2 and 7 respectively.}
\label{fig1}
\end{figure}

\begin{figure}[t]
\includegraphics[scale=0.70]{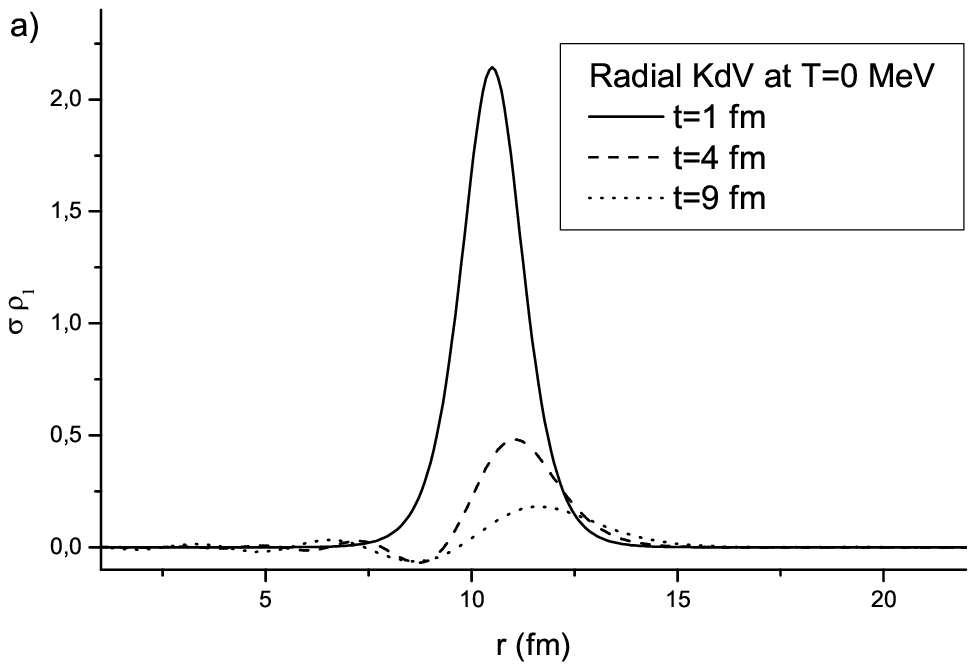}\\
\includegraphics[scale=0.70]{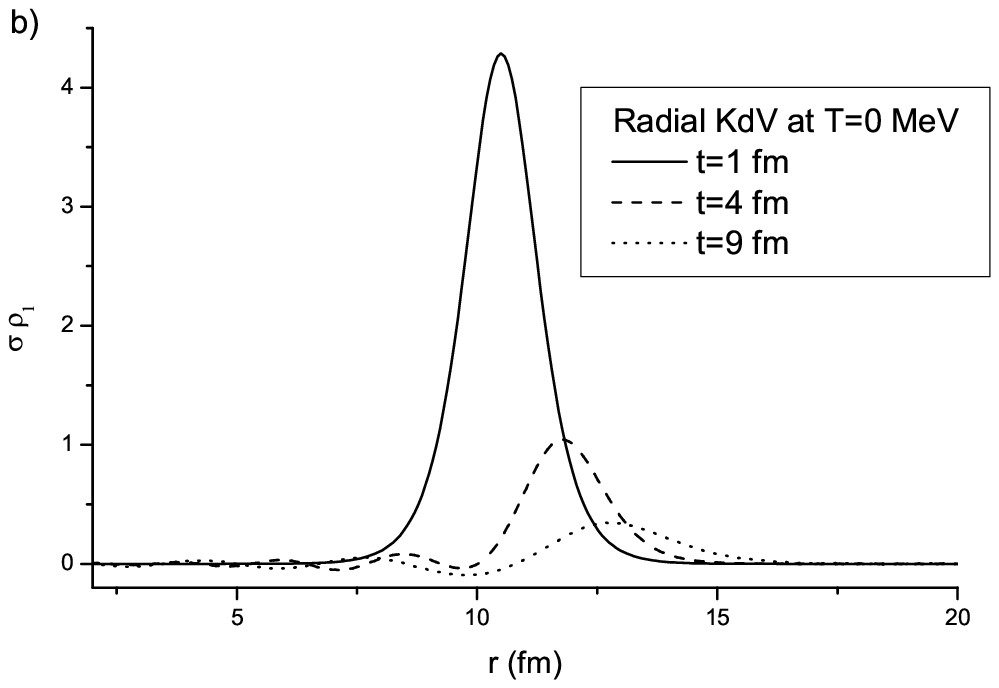}\\
\includegraphics[scale=0.70]{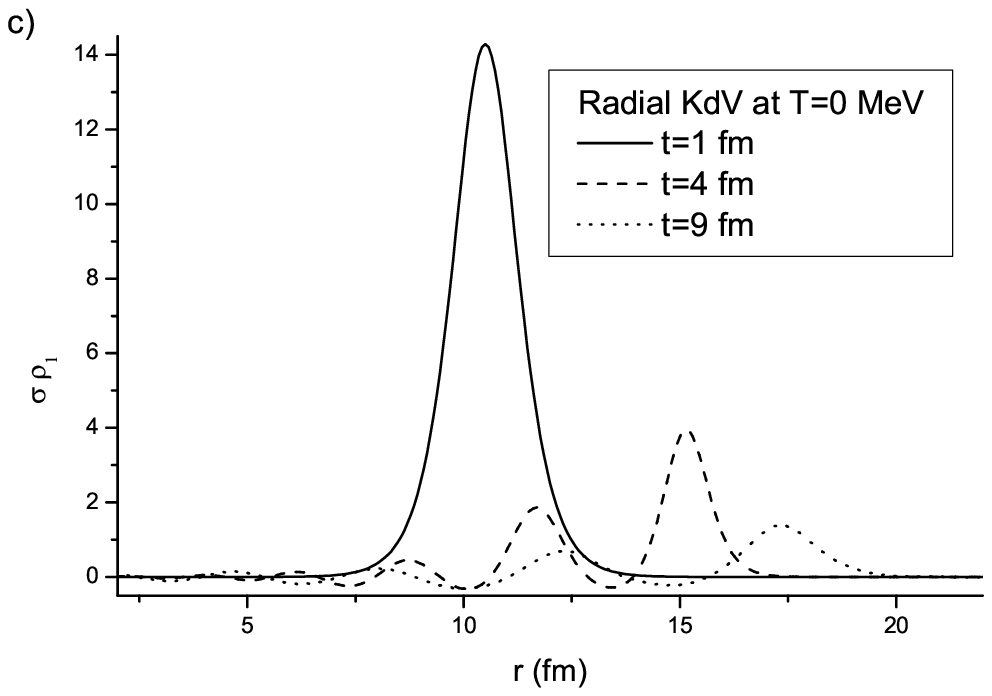}
\caption{The same as Figure 1 for  spherical coordinates.}
\label{fig2}
\end{figure}

\begin{figure}[t]
\includegraphics[scale=0.70]{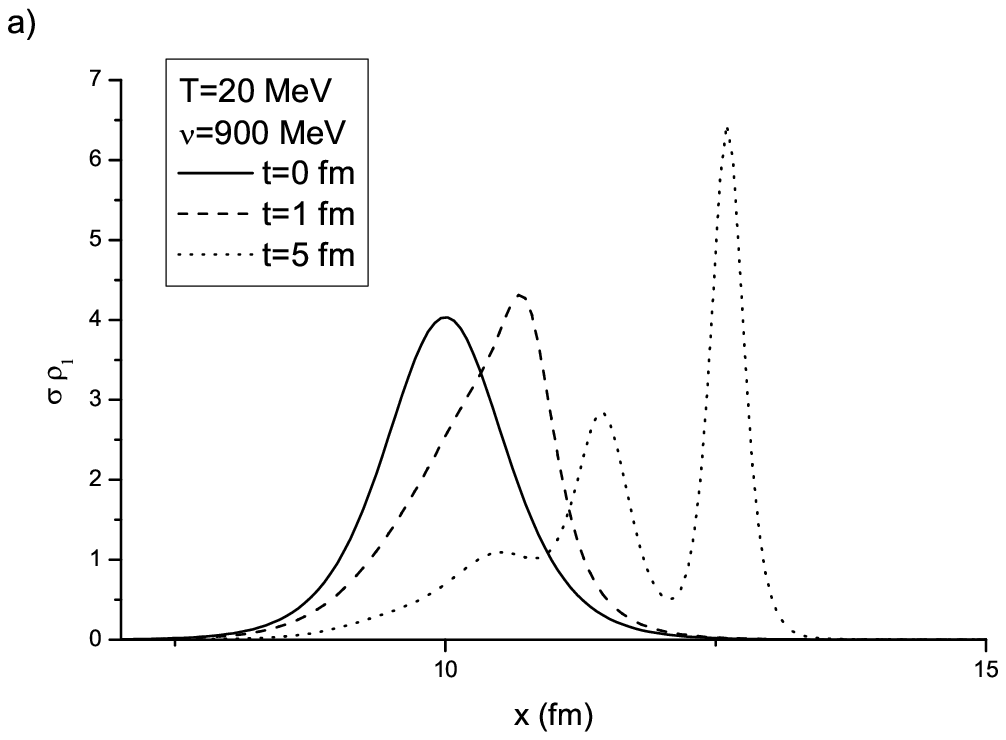}\\
\includegraphics[scale=0.70]{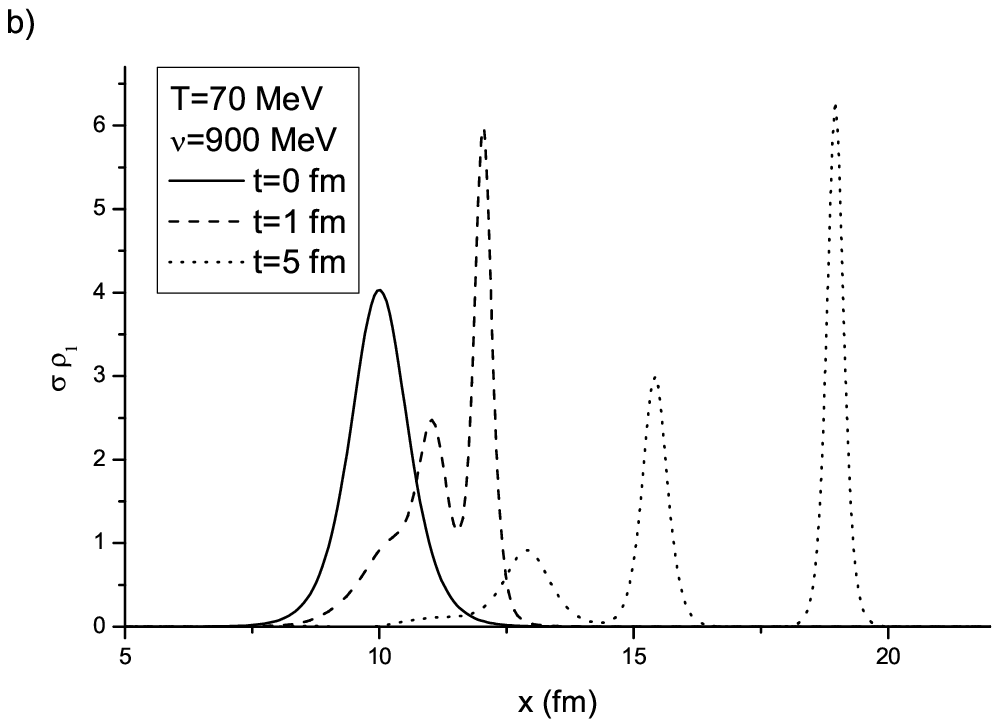}\\
\includegraphics[scale=0.70]{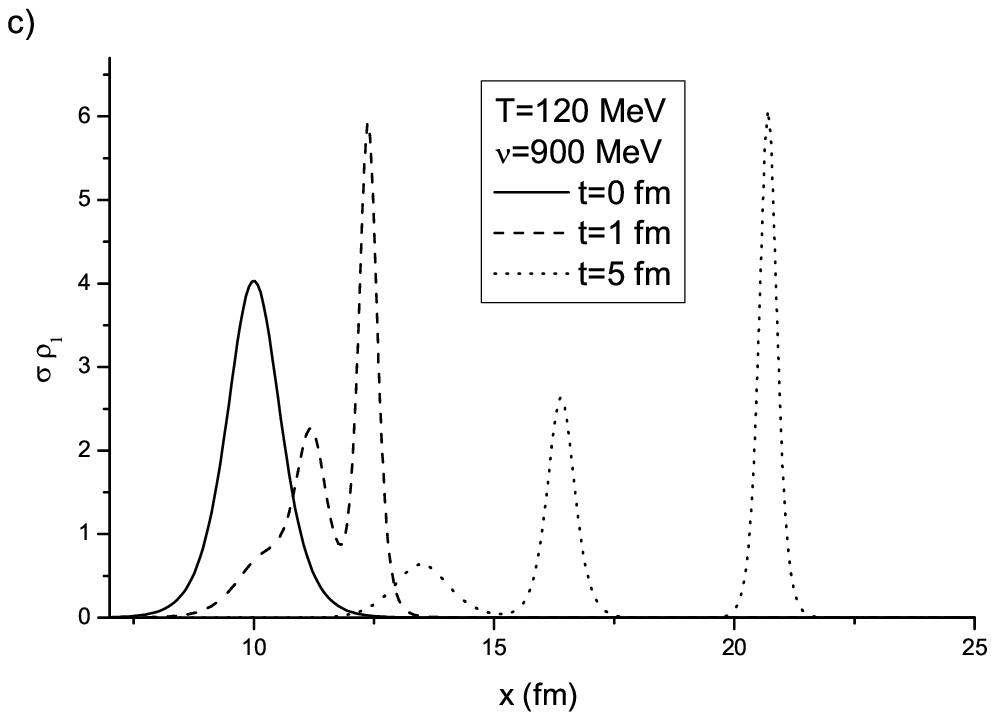}
\caption{Time evolution of  density perturbations in one dimensional Cartesian 
coordinates as a function of the temperature. The pannels show calculations with 
temperatures  $T=20$, $70$ and $120$ MeV respectively.}
\label{fig3}
\end{figure}

\begin{figure}[t]
\includegraphics[scale=0.70]{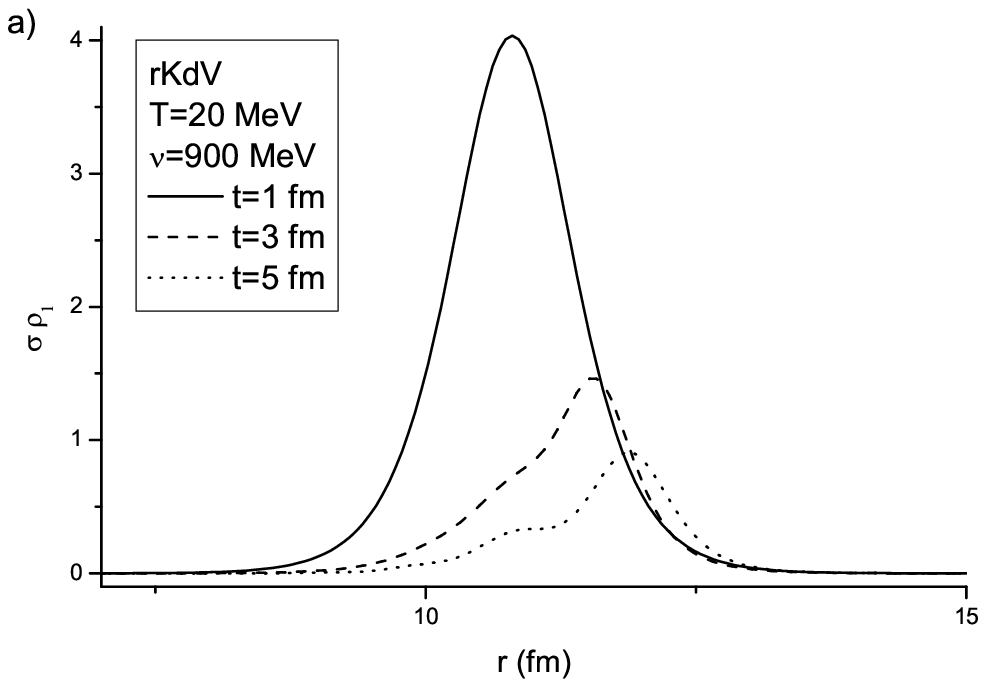}\\
\includegraphics[scale=0.70]{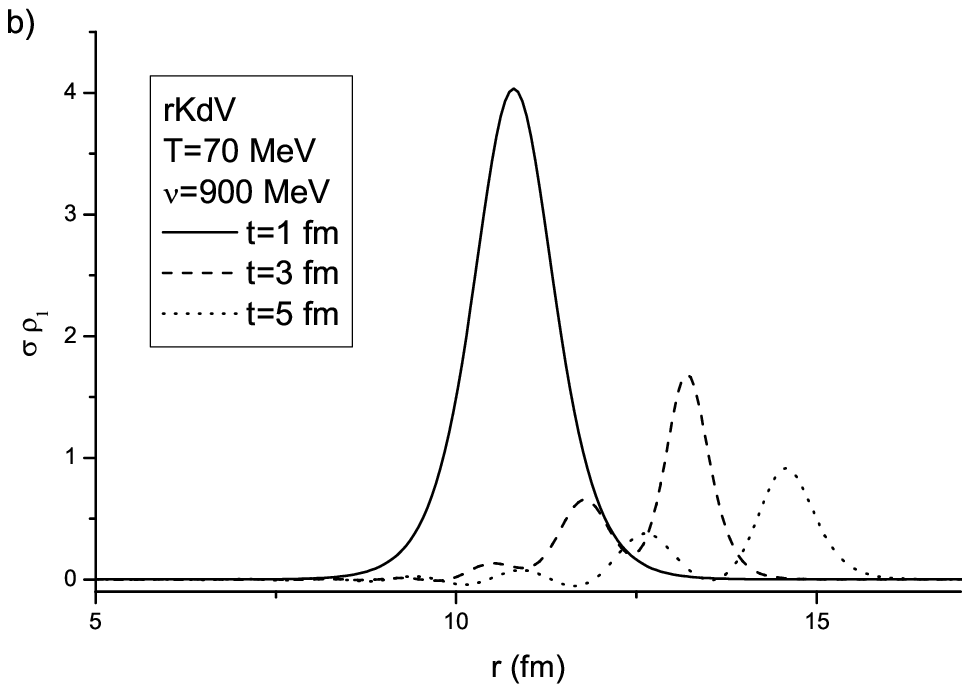}\\
\includegraphics[scale=0.70]{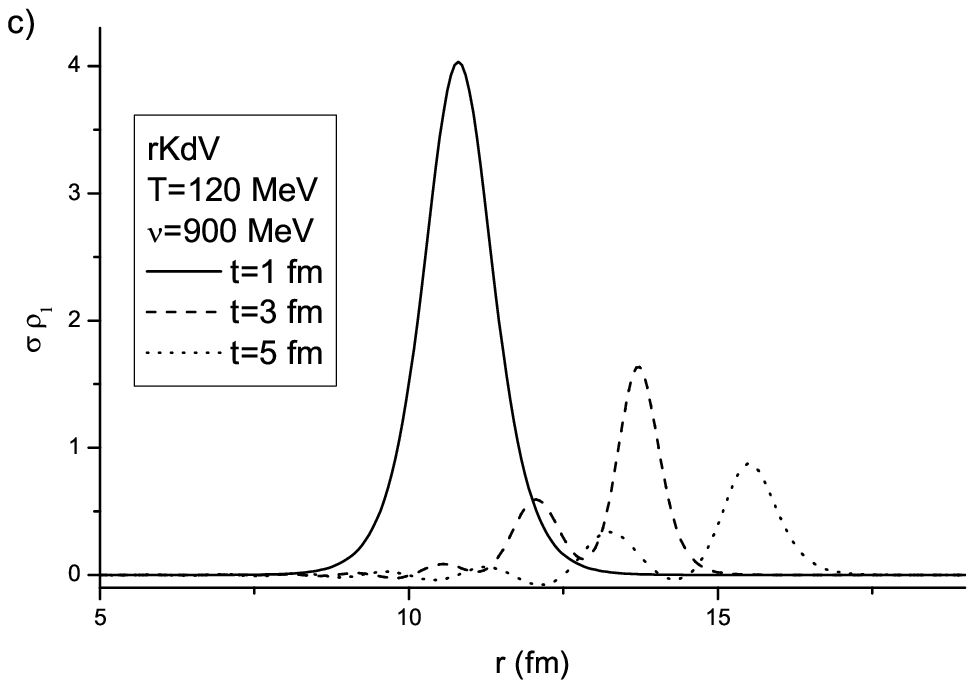}
\caption{The same as Figure 3 for one dimensional  spherical coordinates}
\label{fig4}
\end{figure}


\begin{figure}[t]
\includegraphics[scale=0.70]{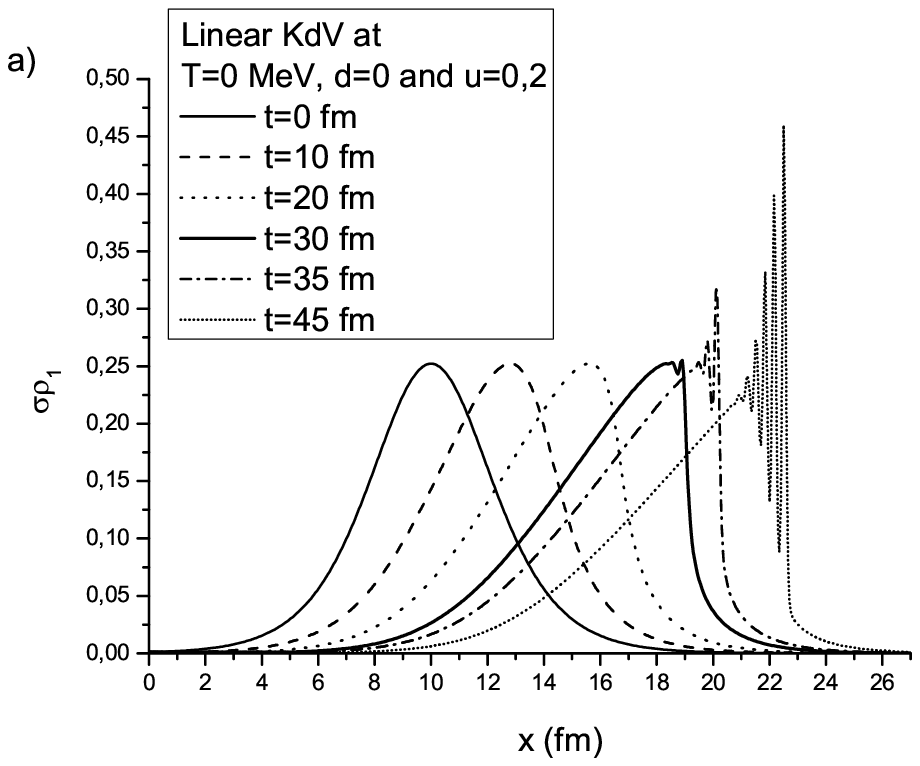}\\
\includegraphics[scale=0.70]{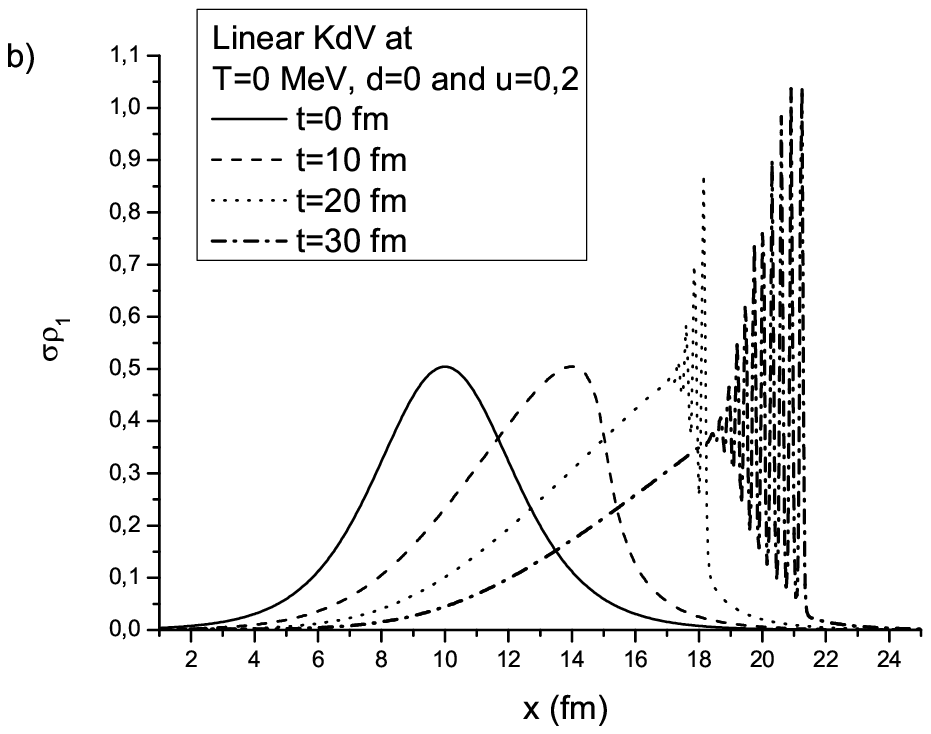}\\
\includegraphics[scale=0.70]{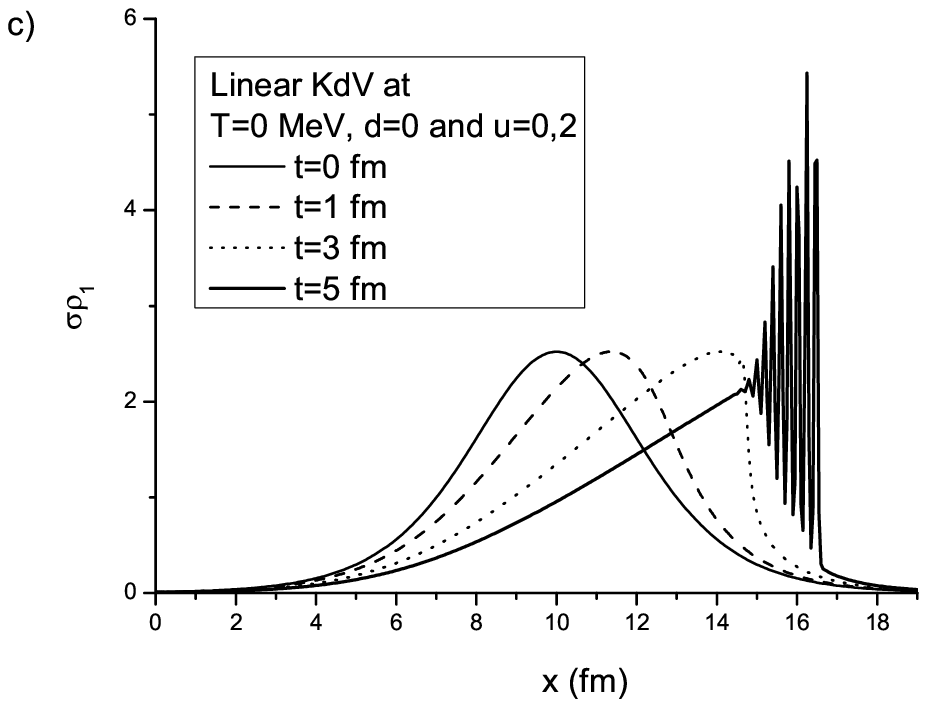}
\caption{Shock wave formation in one dimensional Cartesian coordinates. a) Initial  profile 
with heigth $0.25$ (arbitrary units); b) with heigth $0.50$; c) with heigth $2.5$.}
\label{fig5}
\end{figure}

\begin{figure}[t]
\includegraphics[scale=0.70]{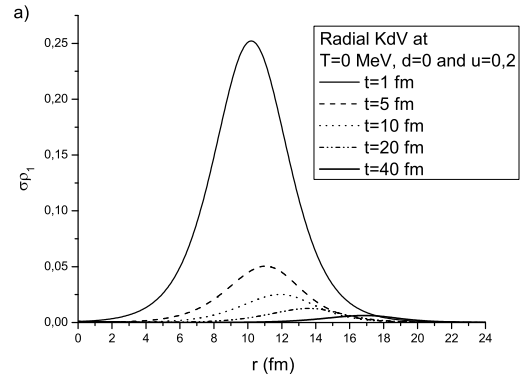}\\
\includegraphics[scale=0.70]{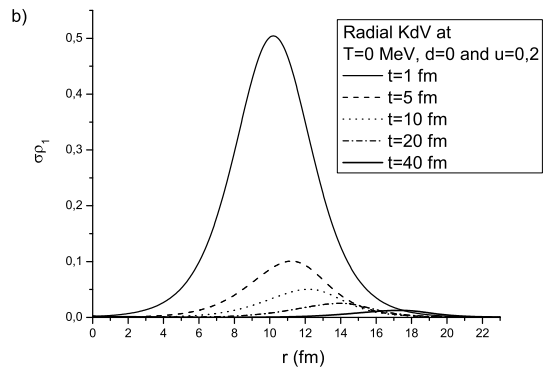}\\
\includegraphics[scale=0.70]{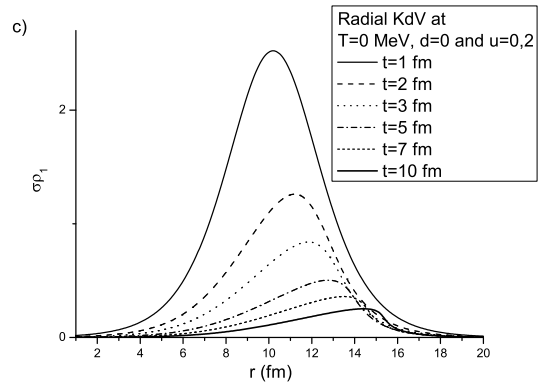}
\caption{The same as Figure  5 in one dimensional spherical coordinates }
\label{fig6}
\end{figure}

\begin{figure}[t]
\includegraphics[scale=0.70]{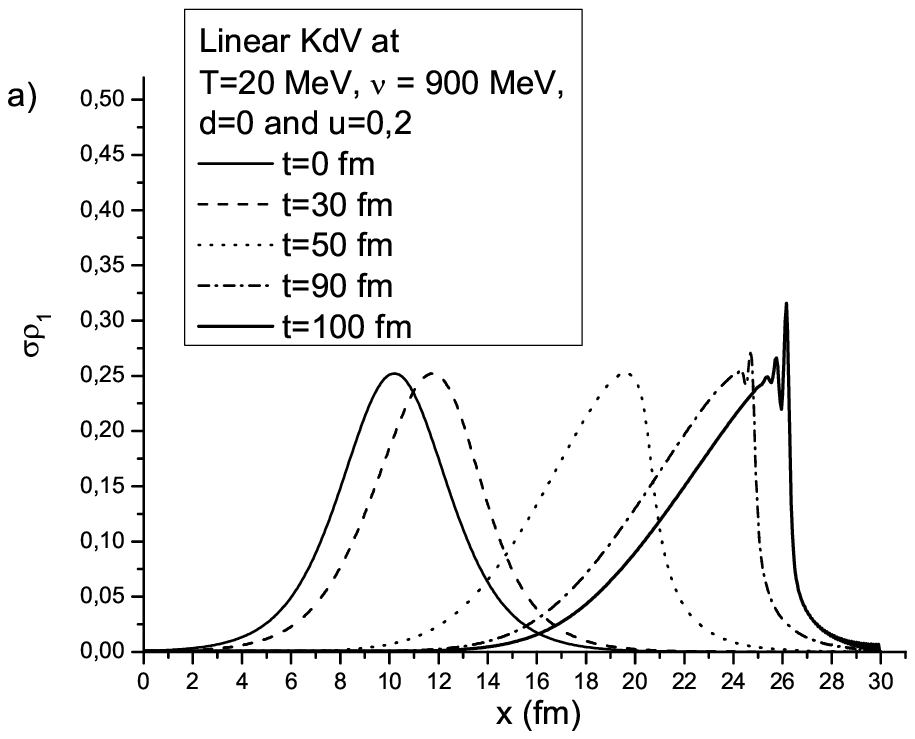}\\
\includegraphics[scale=0.70]{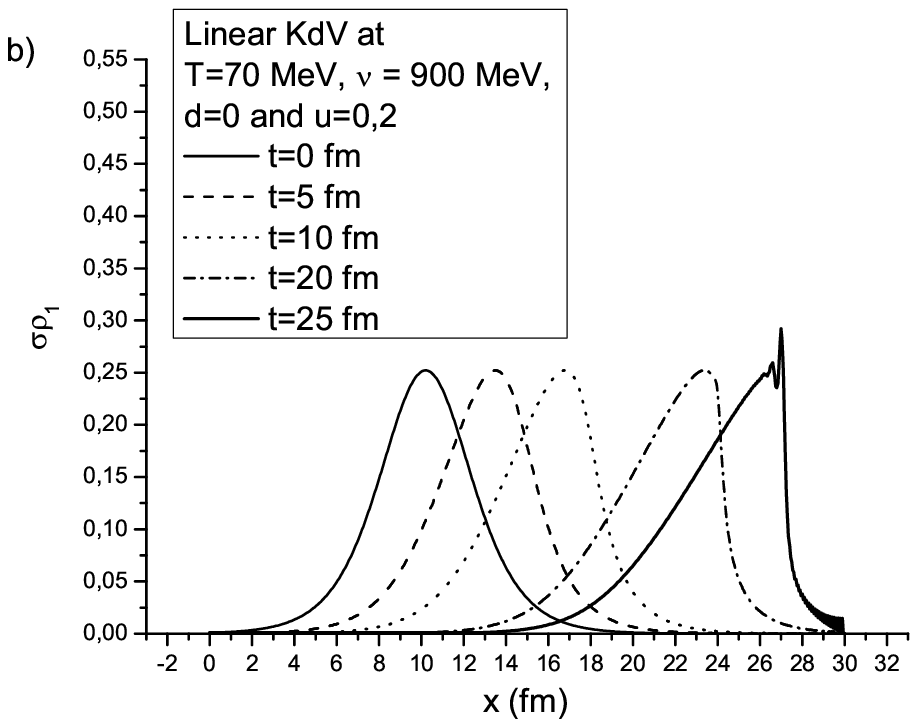}\\
\includegraphics[scale=0.70]{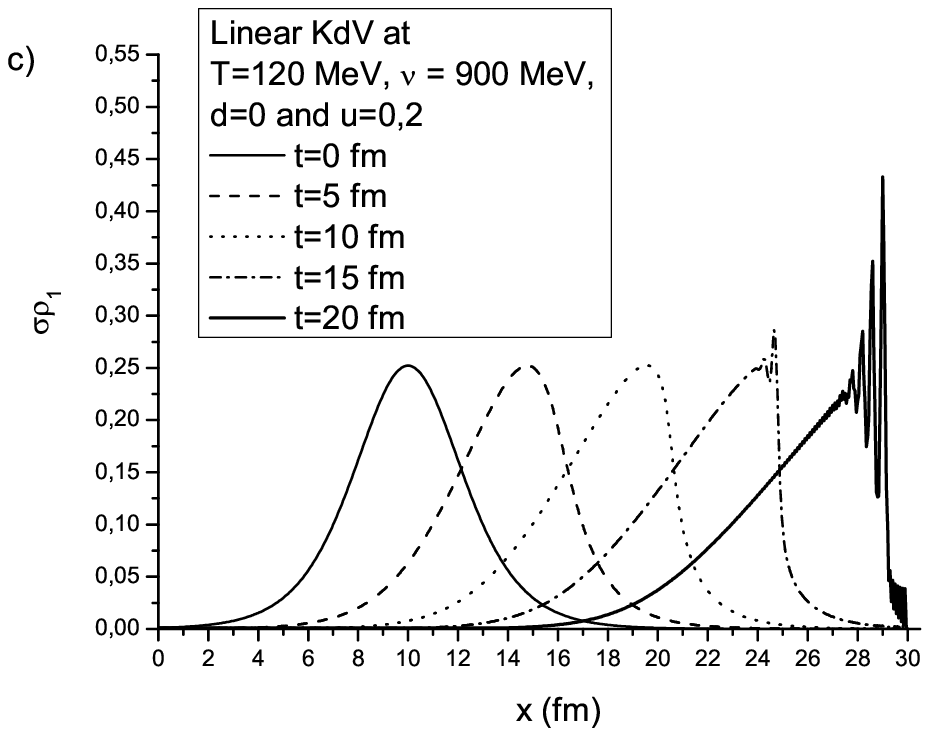}
\caption{Shock wave formation in one dimensional Cartesian 
coordinates for different temperatures.}
\label{fig7}
\end{figure}

\begin{figure}[t]
\includegraphics[scale=0.70]{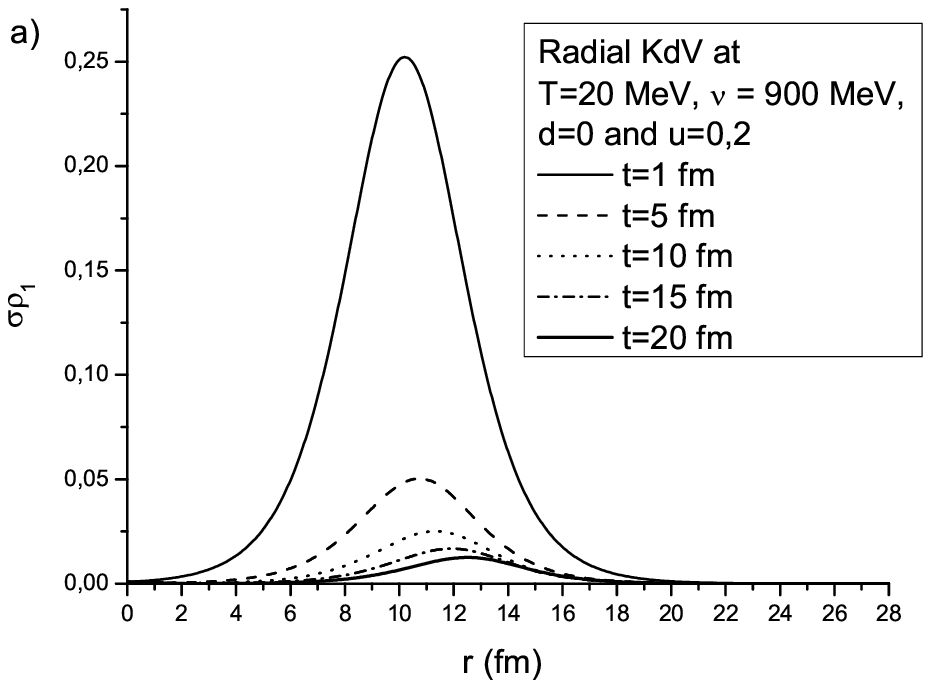}\\
\includegraphics[scale=0.70]{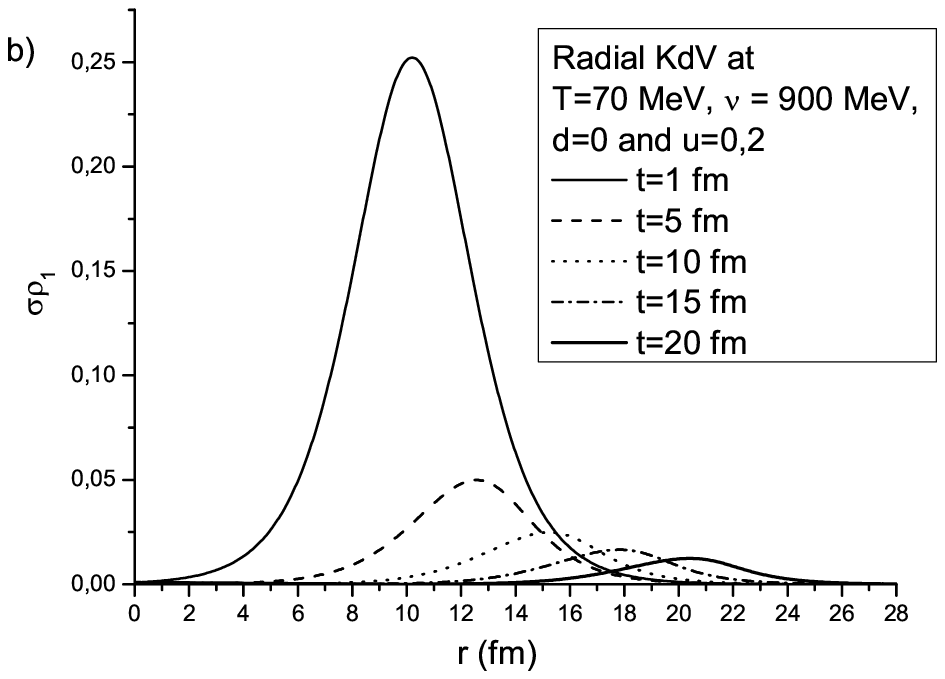}\\
\includegraphics[scale=0.70]{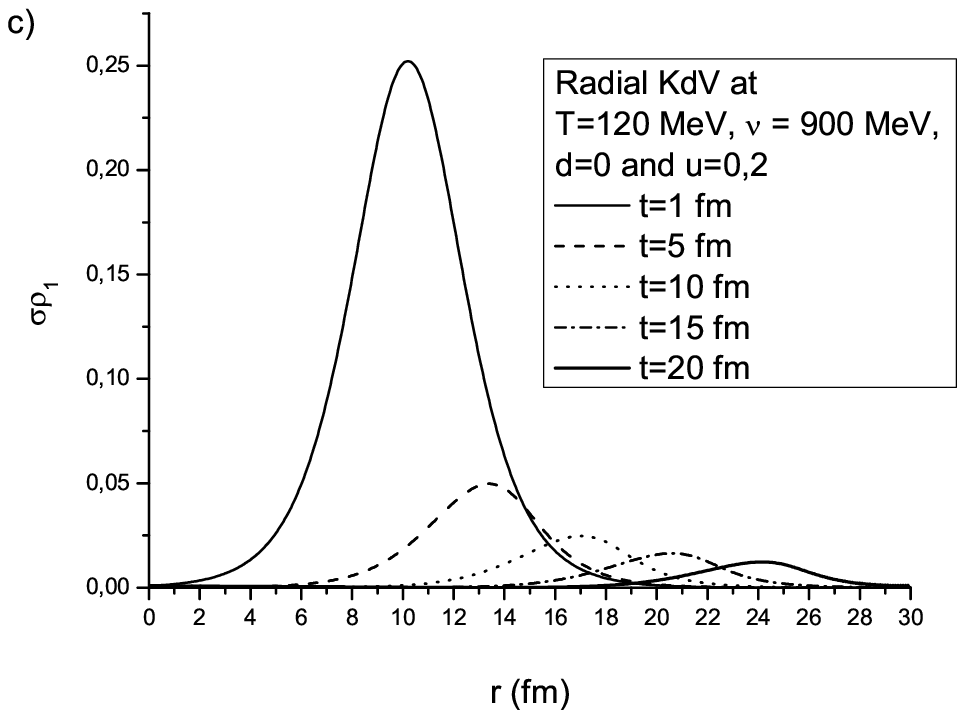}
\caption{The saem as Figure 7 for one dimensional spherical coordinates.}
\label{fig8}
\end{figure}

\subsection{One dimensional Cartesian coordinates}

One dimensional perturbations propagating in cold nuclear matter have been discussed in 
detail in \cite{fn1} and \cite{fn2} and thus we only give here the obtained differential
equation:
\begin{equation}
{\frac{\partial {\hat{\rho}_{1}}}{\partial t}}+
{c_{s}}{\frac{\partial {\hat{\rho}_{1}}}{\partial x}}+
(3-{c_{s}}^{2}){c_{s}}
{\hat{\rho}_{1}}{\frac{\partial{\hat{\rho}_{1}}}{\partial x}}
+\bigg({\frac{{ d \,g_{V}}^{2}\rho_{0}}{2M{{m_{V}}^{4}}{c_{s}}}}\bigg)
{\frac{\partial^{3}{\hat{\rho}_{1}}}{\partial x^{3}}}=0  
\label{lintzero}
\end{equation}
which has the following analytical solution:
\begin{equation}
{\hat{\rho}_{1}}(x,t)={\frac{3(u-{c_{s}})}{{c_{s}}}}(3-{c_{s}}^{2})^{-1}sech^{2}\bigg[
{\frac{{m_{V}}^{2}}{{g_{V}}}}\sqrt{{\frac{(u-{c_{s}}){c_{s}}M}
{2 d \,{\rho_{0}}}}}(x-ut) \bigg] 
\label{lintzerosol}
\end{equation}
At finite temperature we replace  (\ref{epstzero}) by  (\ref{epst}) and use it 
directly in (\ref{068}) without imposing any saturation condition. Everything else is 
the same as described in the last sections. We arrive at the following  equation:
\begin{equation}
{\frac{\partial {\hat{\rho_{1}}}}{\partial t}}+{c_{s}}
{\frac{\partial {\hat{\rho_{1}}}}{\partial x}}
+\bigg(2-{c_{s}}^{2}
-{\frac{{\mu_{B}}{m_{V}}^{2}{c_{s}}^{2}}{2{g_{V}}^{2}{\rho_{0}}}}\bigg){c_{s}}
\hat{\rho_{1}}{\frac{\partial{\hat{\rho_{1}}}}{\partial x}}+
\bigg({\frac{{d \,c_{s}}}{2{{m_{V}}^{2}}}}\bigg)
{\frac{\partial^{3}{\hat{\rho_{1}}}}{\partial x^{3}}}=0
\label{lint}
\end{equation}
with analytical solution given by:
\begin{equation}
{\hat{\rho}_{1}}(x,t)={\frac{3(u-{c_{s}})}{{c_{s}}}}\bigg(2-{c_{s}}^{2}
-{\frac{{\mu_{B}}{m_{V}}^{2}{c_{s}}^{2}}{2{g_{V}}^{2}{\rho_{0}}}}\bigg)^{-1}sech^{2}\bigg[
\sqrt{{\frac{(u-{c_{s}}){m_{V}}^{2}}
{2 d \, c_{s} }}}(x-ut) \bigg] 
\label{lintsol}
\end{equation}

\section{Numerical Results}

The numerical solution of non-linear differential equations is not very difficult, 
but  may be very tricky. We benefited from the reading and hints contained in the 
textbook \cite{dunga}, which has a special section dedicated to solitons.

We start our numerical analysis showing in Fig. \ref{fig1} the solution of 
the linear KdV equation at $T=0$, Eq. (\ref{lintzero}). In the 
upper pannel, Fig. \ref{fig1}a), we use the analytical solution Eq. 
(\ref{lintzerosol}) as initial condition. As expected this pulse 
propagates without dissipation nor dispersion. Any change in the initial 
condition has noticeable consequences. In Fig. \ref{fig1}b), we follow the evolution 
of the numerical solution of  (\ref{lintzero}) for an initial pulse given
by (\ref{lintzerosol}) multiplied by a factor two. As it can be seen, the 
amplitude grows, the width decreases and a second bumps appears propagating  
behind  the first. In Fig. \ref{fig1}c) we start the evolution with  (\ref{lintzerosol}) 
multiplied by a factor seven. Now we have three peaks instead of two. In practical 
cases, the initial conditions  will never
be exactly those needed to generate a single pulse. Therefore, in general we expect 
to see multiple bumps.  In Fig.  \ref{fig1} we can also see that perturbations with higher 
amplitudes propagate faster.

In Fig. \ref{fig2}, we show the equivalent plot for the spherical 
case. In contrast to the linear case there is a strong damping of the pulse. The dependence  
on the initial conditions is also strong. The main peak very rapidly looses height and 
develops secondary bumps. 

In Fig. \ref{fig3} we show for the linear case and for  the 
 ``optimal''  initial condition (\ref{lintzerosol}) the evolution of the 
pulse with time for different temperatures. We can see that, increasing the temperature 
the pulses move faster and go farther. The same feature can be observed in the spherical 
case, as shown in Fig. \ref{fig4}. 

Setting $d=0$ in (\ref{lagr}) we recover the standard non-linear Walecka model. In this 
medium 
the propagation of density perturbations will be governed by the differential equations 
(\ref{esftzero}), (\ref{esft}), (\ref{lintzero}) and  (\ref{lint}) without the terms with 
the 
$d$ factor. The absence of the third order derivative terms leads to a lack of stability of 
the solution. The corresponding differential equations are ``shock wave equations''.  Out of 
smooth initial perturbations these equations create shock waves. We can see this process in 
one dimensional Cartesian coordinates in Fig. \ref{fig5}.  

Figure  5a)  shows  
the solution of Eq.  (\ref{lintzero}) with $d=0$ for the initial condition (\ref{initesf}).
In Figs. 5b) and 5c) the initial profile has been multiplied by a factor 2 and 10 respectively. 
In all cases we observe a steepening of the profile until the formation of the 
shock, followed by the dispersion of the wave. We see that the higher is the initial 
amplitude, the sooner the wave breaking and dispersion occur. 
The solutions of (\ref{esftzero}) with $d=0$ for three different initial profiles can be 
seen in  Fig. \ref{fig6}, where the three pannels, 6a), 6b) and 6c), show the evolution of 
the original profile and then the evolution of this profile multiplied by 2 and by  10 
respectively.  In the spherical case the damping is so strong that  wave breaking hardly 
happens.

In Figure \ref{fig7} we fix one initial profile and study its time evolution for 
three different temperatures. Figs. 7a), 7b) and 7c) show the development of a shock wave 
at $T=20$ MeV, $70$ MeV and $120$ MeV respectively.  As it can be seen, with increasing 
temperatures the pulse moves faster and the shock formation and the subsequent dispersive 
breaking occur sooner.  Fig. \ref{fig8} shows the analogous  plot for spherical coordinates.

\section{Conclusions}

From the equations of relativistic hydrodynamics and with an equation of state obtained from a 
variant of the  non-linear Walecka model we have derived a spherical  KdV-like equation for 
perturbations  in the baryon density. The coefficients of the differential equation are 
determined by the microscopic meson exchange  dynamics. This is an improvement over our previous 
study \cite{fn4}. Moreover we have included temperature effects and solved numerically  the 
resulting differential equations. We have also, for the first time,  obtained numerical 
solutions for the one dimensional Cartesian case at zero and finite temperature. 

The results give a quantitative measure of the dependence of the numerical solutions on the 
initial conditions. We found that, as expected in non-linear problems, the behavior of the 
solutions depends very strongly on the initial conditions. 

The results presented in Fig. \ref{fig3} are a first step towards a realistic study of the 
propagation of a fast leading particle (coming from a jet) crossing hot hadronic matter. They 
suggest that perturbation may propagate for a relatively long distance preserving features of 
the initial peak structure. This is even more true  at higher temperatures.  In contrast, in the 
spherical case shown in Fig. \ref{fig4}, our results show a strong attenuation, indicating that  
localized perturbations will not survive for long distances. They will instead release energy to the medium in a more homogeneous way. This behavior may have consequences for astrophysical 
phenomena and we plan to address this subject in the near future.  Switching off the cubic 
derivative term in the Lagrangian density and recovering the standard non-linear Walecka model, the propagation of initial density pulses generates shock waves, which go through a 
dispersive breaking. Both the propagation and breaking depend strongly on the properties 
(heigth and width) of the initial pulses and on the temperature of the medium. Higher pulses 
move faster and break earlier. The same effect is observed when we increase the temperature. In contrast,   spherical pulses are very insensitive to the initial conditions and to the 
temperature. 

We plan to investigate the consequences of our findings both in 
the relativistic heavy ion physics  and dense stars physics scenarios.

\begin{acknowledgments}
We wish to express our gratitude to S. Duarte, F.O. Dur\~aes, E. Fraga, T. Kodama, 
S. Raha, S. Szpigel, Dou Fu-Quan and A. Gammal  for numerous suggestions and useful 
comments and hints. This work was  partially  financed by the Brazilian funding
agencies CAPES, CNPq and FAPESP. 
\end{acknowledgments}





\begin{thebibliography}{99}


\bibitem{hidro1} J.Y. Ollitrault, arXiv:0708.2433 [nucl-th]; 

\bibitem{hidro2} U. Heinz,   {\sl J. Phys.} {\bf G31}, S717 (2005); 
                for a recent review see P.F. Kolb, U. Heinz, 
                in  ``Quark Gluon Plasma 3'', Editors: R.C. Hwa and X.-N. Wang, World 
                Scientific,  Singapore, (2003) p. 634; nucl-th/0305084.

\bibitem{hidro3} R.B. Clare and D. Strottman, {\sl Phys. Rept.} {\bf 141}, 177 (1986).
 
\bibitem{hidro4}  Y.~Hama, T. Kodama and O. Socolowski Jr., 
                {\sl Braz.  J.  Phys.} {\bf 35}, 24 (2005); 
                Y.~Hama and F.S.~Navarra,  {\sl Phys. Lett.} {\bf B129}, 251 (1983);  
                {\sl Z. Phys.} {\bf C53}, 501 (1992).


\bibitem{v2}  K.~H.~Ackermann {\it et al.}  [STAR Collaboration],
  Phys.\ Rev.\ Lett.\  {\bf 86}, 402 (2001); 
  P.~Huovinen, P.~F.~Kolb, U.~W.~Heinz, P.~V.~Ruuskanen and S.~A.~Voloshin,
  Phys.\ Lett.\   {\bf B503}, 58 (2001); 
  S.~S.~Adler {\it et al.}  [PHENIX Collaboration],
  Phys.\ Rev.\ Lett.\  {\bf 91}, 182301 (2003). 





\bibitem{shock}  L.~M.~Satarov, H.~Stoecker and I.~N.~Mishustin,
  Phys.\ Lett.\   {\bf B627}, 64 (2005);
  T.~Renk and J.~Ruppert,
  Phys.\ Rev.\   {\bf C73}, 011901 (2006);
  T.~Renk and J.~Ruppert,
  Phys.\ Lett.\   {\bf B646}, 19 (2007).


\bibitem{away1}  S.~S.~Adler {\it et al.}  [PHENIX Collaboration],
  Phys.\ Rev.\ Lett.\  {\bf 97}, 052301 (2006).

\bibitem{away2} J. Adams et al., STAR Collab.  {\sl Phys. Rev. Lett. } {\bf 95},
               152301 (2005).

\bibitem{frsw} G.N. Fowler, S. Raha, N. Stelte and R.M. Weiner, 
               {\sl Phys. Lett.} {\bf B115}, 286 (1982); 
               S. Raha and R.M. Weiner, 
               {\sl Phys. Rev. Lett.} {\bf 50}, 407 (1983);  
               E.F. Hefter, S. Raha and R.M. Weiner, 
               {\sl Phys. Rev.} {\bf C32}, 2201 (1985).

\bibitem{rww}  S. Raha, K. Wehrberger and R.M. Weiner, 
               {\sl Nucl. Phys.} {\bf A433}, 427 (1984).

\bibitem{abu} A.Y. Abul-Magd, I. El-Taher and F.M. Khaliel, 
              {\sl Phys. Rev.} {\bf C45}, 448 (1992).


\bibitem{lala} G.A. Lalazissis, J. K\"onig and P. Ring, {\sl Phys. Rev.} {\bf C55}, 
               540 (1997).


\bibitem{furn} R.J. Furnstahl, {\sl Lect. Notes Phys.} {\bf 641}, 1 (2004);
               B.D. Serot,  {\sl Int. J. Mod. Phys.} {\bf A19S1}, 107 (2004)  
                     and  references therein.

\bibitem{serot} B.D. Serot and J.D. Walecka, 
                {\sl Advances in Nuclear Physics} {\bf 16}, 1 (1986).



\bibitem{fn1} D.A. Foga\c{c}a and  F.S. Navarra, 
               {\sl Phys. Lett. } {\bf B639}, 629 (2006).

\bibitem{fn2} D.A. Foga\c{c}a and  F.S. Navarra, 
               {\sl Phys. Lett. } {\bf B645}, 408 (2007).

\bibitem{fn3} D.A. Foga\c{c}a and  F.S. Navarra, 
  {\sl Nucl. Phys.} {\bf A790}, 619c (2007).

\bibitem{fn4} D.A. Foga\c{c}a and  F.S. Navarra, 
             {\sl  Int. J. Mod. Phys.}   {\bf E 16}, 3019 (2007).


\bibitem{xue} J.-K. Xue,  {\sl Phys. Lett. } {\bf A322}, 225 (2004).


\bibitem{pu} L.K. Pu, D. Fu-Quan, S. Jian-An, D. Wen-Shan, S. Yu-Ren,
             {\sl Chinese Phys.} {\bf 14}, 33   (2005); 
             {\sl Chinese Phys. Lett.} {\bf  18}, 1088 (2001).



\bibitem{wein} S. Weinberg,``Gravitation and Cosmology'', New York: Wiley, 1972.


\bibitem{land} L. Landau and  E. Lifchitz, ``Fluid Mechanics'',  
                Pergamon Press, Oxford, (1987).



\bibitem{elze} H.-T. Elze, Y. Hama, T. Kodama, M. Makler and J. Rafelski,
		         {\sl J. Phys. G: Nucl. Part. Phys.} {\bf 25}, 1935 (1999).


\bibitem{reif} R. Reif, ``Fundamentals of statistical and thermal physics'', New York:
               McGraw-Hill, 1965.


\bibitem{amigos} For a recent example see  
M.~Chiapparini, M.~E.~Bracco, A.~Delfino, M.~Malheiro, D.~P.~Menezes and C.~Provid\^encia,
arXiv:0711.3631 [hep-ph] and references therein.


                                                        
\bibitem{shin} N. Sasaki, O. Miyamura, S. Muroya, C. Nonaka, 
               {\sl Europhys.  Lett.}  {\bf 54}, 38 (2001); 
               {\sl Phys. Rev.} {\bf C62}, 011901 (2000). 




\bibitem{davidson} R.C. Davidson, ``Methods in Nonlinear Plasma Theory'', 
                   Academic Press, New York an London, 1972.





\bibitem{dunga} R.H. Landau and M.J. Paez Mejia,    ``Computational physics: problem 
                solving with computers'',  New York: John Wiley, 1997. 

 
\end{thebibliography}
\end{document}